\documentclass[runningheads]{llncs}
\usepackage{graphicx}
\usepackage{lineno}
\usepackage{tabularx, multirow, multicol}
\usepackage{array}
\makeatletter
\newcommand{\thickhline}{%
	\noalign {\ifnum 0=`}\fi \hrule height 1pt
	\futurelet \reserved@a \@xhline
}
\newcolumntype{"}{@{\hskip\tabcolsep\vrule width 1pt\hskip\tabcolsep}}
\makeatother

\usepackage{amsmath}
\usepackage{algorithm}
\usepackage[noend]{algpseudocode}
\usepackage{hyperref} 
\algrenewcommand\algorithmicindent{0.4em} 
\algrenewcommand\textproc{}

\usepackage{enumitem} 
\usepackage{caption}
\usepackage{subcaption}

\usepackage[dvipsnames,svgnames,x11names]{xcolor}
\usepackage[final]{changes}
\definechangesauthor[color=Red]{song}
\definechangesauthor[color=NavyBlue]{B} 
\definechangesauthor[color=DarkGreen]{fu} 

\begin{document}
\title{Designing a 3D Parallel Memory-Aware Lattice Boltzmann Algorithm on Manycore Systems}

\titlerunning{Designing a 3D Parallel Memory-Aware LBM Algorithm on Manycore}
%
\author{Yuankun Fu\inst{1}\orcidID{0000-0003-2453-9310}\and 
Fengguang Song\inst{2}\orcidID{0000-0001-7382-093X}} 
%
%
\institute{Purdue University, Indianapolis, IN, \email{fu121@purdue.edu} \and
Indiana University-Purdue University, Indianapolis, IN, \email{fgsong@cs.iupui.edu}}
\maketitle              

\begin{abstract}
Lattice Boltzmann method (LBM) is a promising approach to solving Computational Fluid Dynamics (CFD) problems, however, its nature of memory-boundness limits nearly all LBM algorithms' performance
on modern computer architectures.
This paper introduces novel sequential and parallel 3D memory-aware LBM algorithms to 
optimize its memory access performance. 
The introduced new algorithms combine the features of single-copy distribution, single sweep, swap algorithm, prism traversal, and merging two 
temporal time steps.
We also design a parallel methodology to guarantee thread safety and reduce synchronizations in the parallel LBM algorithm.
At last, we evaluate their performances on three high-end manycore systems and demonstrate that our new 3D memory-aware LBM algorithms outperform the state-of-the-art \added[]{Palabos software (which realizes the Fuse Swap Prism LBM solver)} by up to 89\%.

\end{abstract}
%
%
%


\section{Introduction}
\label{sec:intro-motivation}

Computational Fluid Dynamics (CFD) simulations 
have revolutionized the design process in various scientific, engineering, industrial, and medical fields.
The current Reynolds averaged Navier-Stokes (RANS) methods can solve steady viscous transonic and supersonic flows, but are not able to reliably predict turbulent separated flows \cite{witherden2017future}.
Lattice Boltzmann method (LBM) is a young and evolving approach to solving these problems in the CFD community \cite{coreixas2019comprehensive}. 
It originates from a mesoscale description of the fluid (based on the Boltzmann equation), and directly incorporates physical terms 
to represent complex physical phenomena, such as multi-phase flows, reactive and suspension flows, etc.
Besides, many {\it collision models} have been developed for LBM to improve its stability to the second order of numerical accuracy when simulating high Reynolds number flows~\cite{coreixas2019comprehensive}. 

\added[]{However, it is challenging to achieve high performance for LBM algorithms, since LBM has large data storage costs and is highly memory-bound on current architectures \cite{succi2019towards}.}
Driven by our prior work \cite{fu2018designing} to merge multiple collision-streaming cycles (or time steps) in 2D,
this study aims to augment the memory-awareness idea to support parallel 3D LBM to optimize data re-utilization.
Although it might seem to be straightforward to move from the 2D space to 3D space, 
it is significantly much more difficult to design an efficient 3D memory-aware LBM algorithm.
In this paper, we target solving the following three main challenges. 
(1) As geometries change from 2D to 3D, the required data storage increases from $O(N^2)$ to $O(N^3)$,
meanwhile data dependencies of the lattice model becomes much more complicated. 
There exist single-copy distribution methods to reduce data storage cost by half, but they require following a particular traversal order. 
Can we combine the best single-copy distribution method with our idea of merging multiple collision-streaming cycles to design a 3D memory-aware LBM with higher performance?
(2) If the combination is possible, since normal 3D tiling~\cite{rivera2000tiling} does not apply to this case, how to additionally explore the spatial locality? 
(3) When designing the parallel 3D memory-aware LBM, 
a non-trivial interaction occurs at the boundaries between threads,
how to guarantee thread safety and avoid race conditions?
Although some existing works use wavefront parallelism to explore the temporal locality, they insert frequent layer-wise synchronizations among threads every time step \cite{liu2017accelerating,wellein2009efficient}. In this paper, we aim to reduce the synchronization cost among parallel threads.

To the best of our knowledge, this paper makes the following contributions.
First, we design both sequential and parallel 3D memory-aware LBM 
algorithms that combine five features: single-copy distribution, loop fusion (single sweep), swap algorithm, prism traversal, and merging two collision-streaming cycles.
Second, we present a parallelization method to keep the thread safety on the intersection layers among threads and reduce the synchronization cost in parallel.
At last, two groups of experiments are conducted on three different manycore architectures, followed by performance analysis.
The first group of sequential experiments (i.e., using a single CPU core) shows that our memory-aware LBM outperforms the state-of-the-art \added[]{Palabos (Fuse Swap Prism LBM solver)}\cite{Palabos} by up to 19\% on a Haswell CPU and 15\% on a Skylake CPU. 
The second group evaluates the performance of parallel algorithms.
The experimental results show that our parallel 3D memory-aware LBM outperforms \added[]{Palabos} by up to 89\% on a Haswell node with 28 cores, 
85\% on a Skylake node with 48 cores, 
and 39\% on a Knight Landing node with 68 cores.

\section{Related Work}
\label{sec:LBM-relate}

Existing research on designing efficient LBM algorithms mainly focuses on optimizing memory accesses within one time step of LBM due to its iterative nature.
For instance, a few LBM algorithms (e.g., swap~\cite{LBM-swap,valero2017reducing}, AA~\cite{LBM-AA}, shift~\cite{LBM-shift}, and
esoteric twist~\cite{LBM-esoteric-twist},
etc.) retain a single copy of the particle distribution data \added[]{(i.e., ``single-copy distribution'')}, 
and optimize the memory access pattern in the LBM streaming kernel, but each of the algorithms needs to follow a set of constraints 
\added[id=fu, comment={Now We have more space to insert all these contents into the paper.}]{(e.g., swap requires predefined order of discrete cell velocities~\cite{latt2007technical}, AA requires distinguishing between even and odd time steps, shift requires extra storage~\cite{latt2007technical}, esoteric twist requires only one version of the LB kernel~\cite{wittmann2013comparison}, etc.)} 
\cite{vardhan2019moment} uses a moment-based representation with extra distribution pseudo domain to further reduce the storage cost.  
Some works hide the inter-process communication cost on multicore accelerators \cite{crimi2013early}, 
and achieve large-scale parallelization on HPC systems \cite{randles2013performance} and GPU \cite{LBM-AA}. 
\added[id=fu]{\cite{zeiser2008introducing} introduces a cache oblivious blocking 3D LBM algorithm, 
but it has an irregular parallelism scheme due to its recursive algorithm design.}
In summary, the above methods focus on optimizations within one time step.
\added[]{Differently, our 3D memory-aware LBM aims to adopt the efficient single-copy distribution scheme, and 
design new methodologies to merge two collision-streaming cycles to explore both temporal and spatial data locality at the same time for achieving higher performance.} 
Another category of works manages to accelerate LBM by 
wavefront parallelism,
which generally groups many threads to successively compute on the same spatial domain. 
\cite{liu2017accelerating} presents a shared-memory wavefront 2D LBM together with loop fusion, loop bump, loop skewing, loop tiling, and semaphore operations.
But due to its high synchronization cost incurred by many implicit barriers in wavefront parallelism,
their parallel performance has only 10\% of speedup on average.
\cite{habich2009enabling} presents a shared-memory wavefront 3D LBM with two-copy distributions, 
and does not use spatial locality techniques such as loop fusion and loop blocking. 
\cite{wellein2009efficient} presents a shared-memory wavefront 3D Jacobi approach together with spatial blocking. 
\added[]{It uses two-copy distributions and has simpler 6-neighbors dependencies (rather than the 19 or 27 neighbors in 3D LBM).}
\added[id=fu]{\cite{malas2015multicore} combines the wavefront parallelism with diamond tiling.}
By contrast, our 3D memory-aware LBM does not use the wavefront parallelism, \added[]{but judiciously} contains three light-weight synchronization barriers every two collision-streaming cycles.
\added[]{In addition,} we partition the simulation domain and assign a local sub-domain to every thread, 
rather than all threads work on the same sub-domain in wavefront parallelism.
\added[]{In each sub-domain, each thread in our algorithm computes multiple time steps at once, 
rather than one thread computes one time step at a time in wavefront parallelism.}
In addition, \added[]{each of our threads also utilizes prism techniques to optimize spatial locality.}
This strategy in particular favors new manycore architectures, which tend to have increasingly larger cache sizes. 

\added[]{Modern parallel software packages that support LBM can be classified into two categories based upon their underlying data structures.}
One category adopts matrix-based memory alignment at the cell level (e.g., Palabos~\cite{latt2020palabos}, OpenLB~\cite{heuveline2007openlb}, HemeLB~\cite{Hemelb-paper}, HemoCell~\cite{Hemocell-paper}).  
Since neighbors can be easily found through simple index arithmetics in this case, they are \added[]{more suitable for} simulations with dense geometries. 
The other category adopts adjacent list data structures \added[]{(e.g.,} Musubi~\cite{Musubi}, waLBerla~\cite{waLBerla-paper}, HARVEY~\cite{randles2013performance}).
They are \added[]{often used for} simulating \added[]{domains} with sparse and irregular geometries, \added[]{but their} cells require 
\added[]{additional} memory of pointers, and double the memory \added[]{consumption} in the worst case.
\added[id=fu]{In this study, we choose the widely-used and efficient matrix-based data structure in the LBM community, and select the state-of-the-art Palabos library as the baseline,
since Palabos 
provides a broad modeling framework, supports applications with complex physics, and shows high computational performance.} 

\added[id=fu]{\cite{perepelkina2018lrnla} 
designs a locally recursive non-locally asynchronous (LRnLA) conefold LBM algorithm,
which uses recursive Z-curve arrays for data storage, and recursively subdivides the space-time dependency graph into polytopes to update lattice nodes.
However, our work 
uses a more directly accessible matrix-based data storage and has a regular memory access pattern. 
Besides, our prism traversal 
can independently or integrate with merging two time steps to operate on the lattice nodes, 
while \cite{perepelkina2018lrnla} operates on the dependency graph.}



    
    

\section{Baseline 3D LBM Algorithm}
\label{sec:3D-fundamental}

The baseline 3D LBM algorithm in this paper is called \textit{Fuse Swap LBM} as shown in Alg.\ref{alg:3D-fuse}, which involves three features: single-copy distribution, swap algorithm, and loop fusion.
We choose the swap algorithm~\cite{LBM-swap} since it is relatively simpler than the other single-copy distribution methods, 
and is more efficient to use simple index arithmetic to access neighbors in the matrix-based memory organization. 
The swap algorithm replaces the copy operations between a cell and its neighbors in the streaming kernel by a  value swap, thereby it is in-place and does not require the second copy.
But when combining it with loop fusion, 
we must guarantee that the populations of neighbors involved in the swap are already in a post-collision state to keep thread safety~\cite{latt2007technical}.

\begin{figure}[h!]
     \centering
     \begin{subfigure}[b]{0.27\textwidth}
         \centering
         \includegraphics[width=\textwidth]{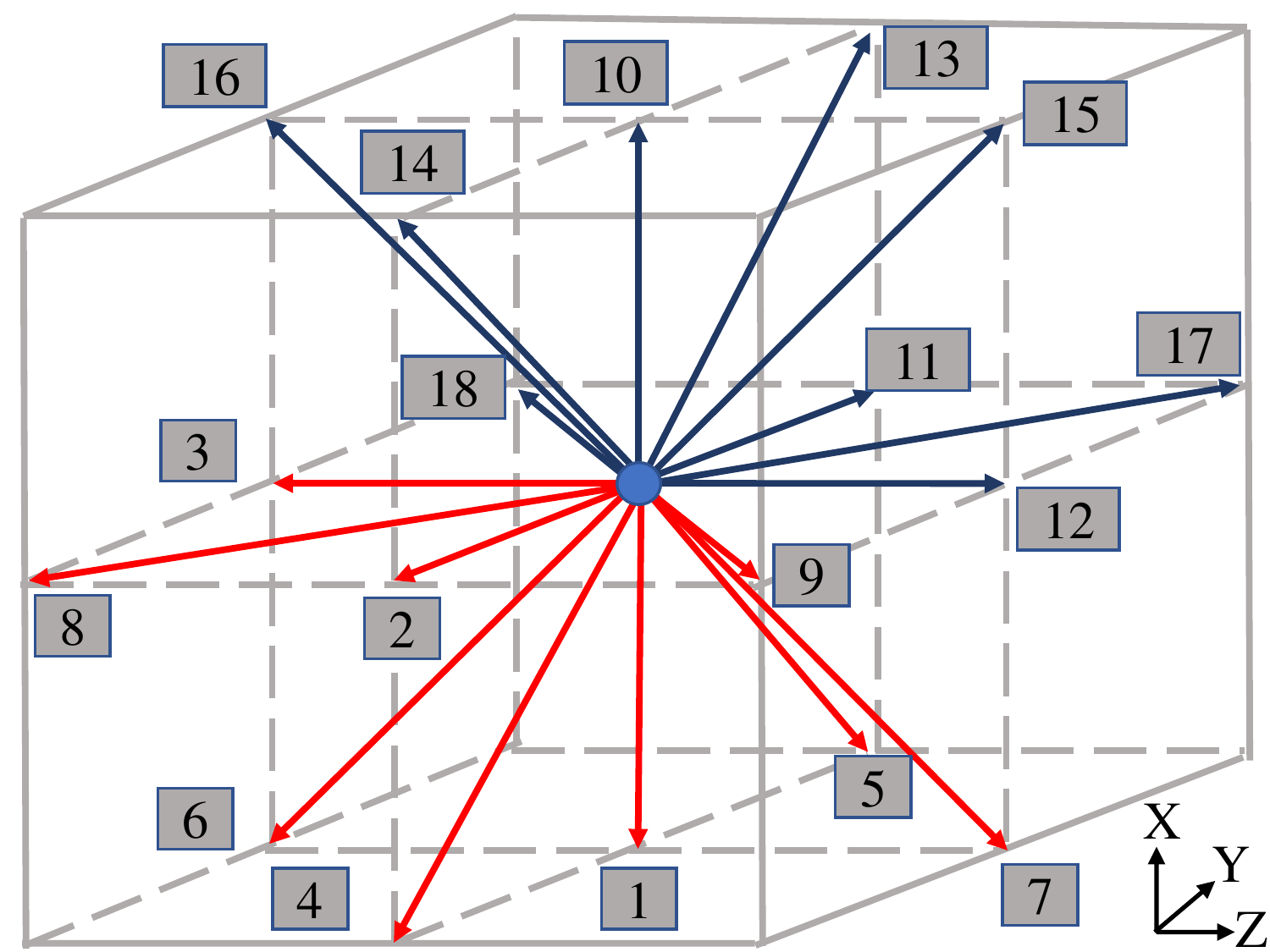}
         \caption{\small $swap\_stream$} 
         \label{fig:3D-swap-stream}
     \end{subfigure}
     \qquad
     \begin{subfigure}[b]{0.27\textwidth}
         \centering
         \includegraphics[width=\textwidth]{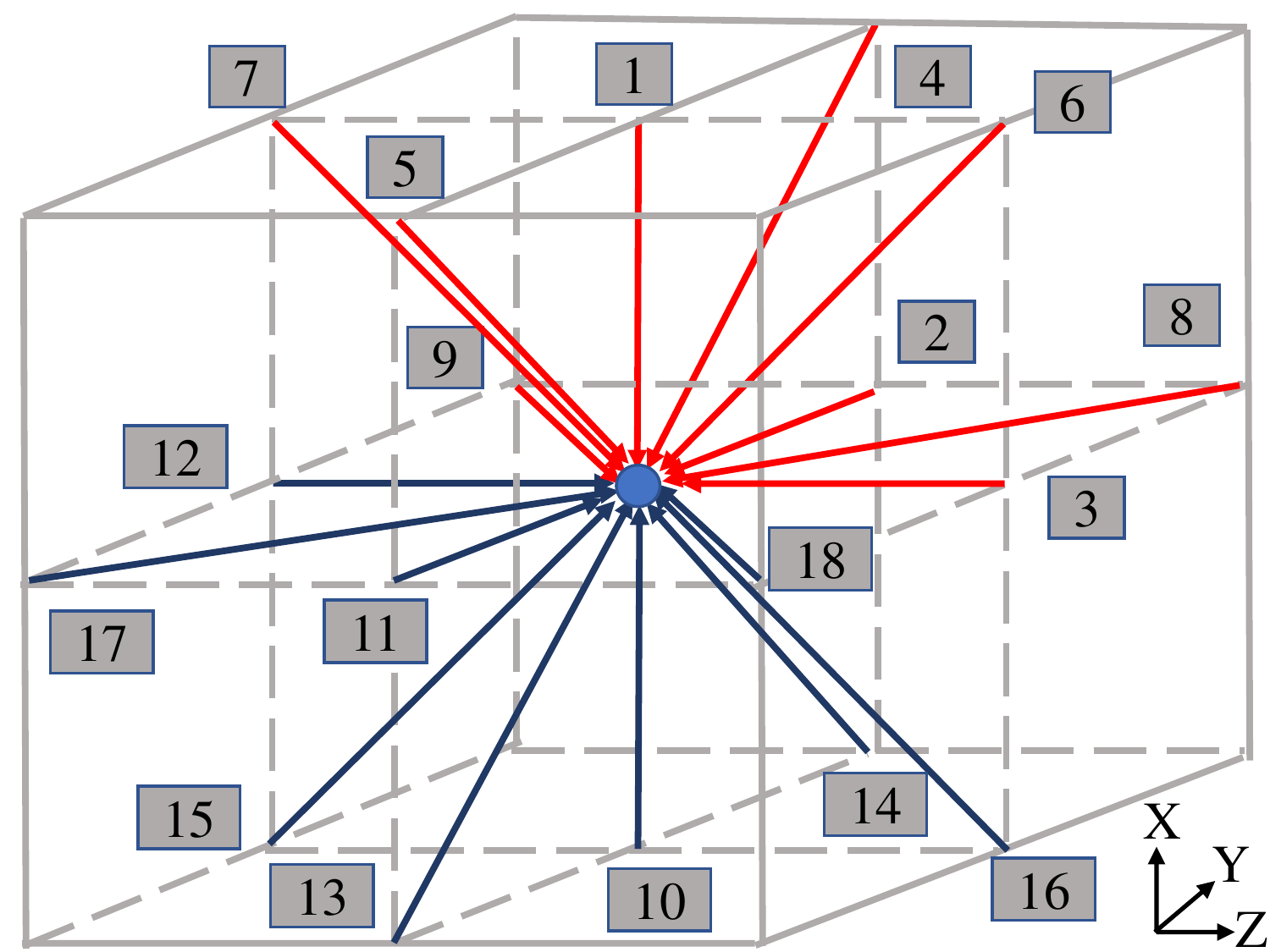}
         \caption{\small $revert$} 
         \label{fig:3D-collide-revert}
     \end{subfigure}
     \qquad
     \begin{subfigure}[b]{0.27\textwidth}
         \centering
         \includegraphics[width=\textwidth]{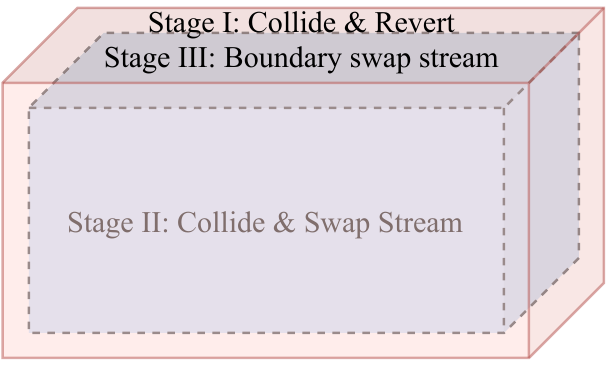}
         \caption{\small Three stages computation.}
         \label{fig:3D-fuse}
     \end{subfigure}
        \caption{Two operations and three stages computation used in sequential 3D Fuse Swap LBM.}
        \label{fig:3D-swap-stream-revert}
\end{figure}

The work-around solution is to adjust the traversal order of simulation domains with a predefined order of discrete cell velocities~\cite{latt2007technical}. 
Thus each cell can stream its post-collision data by swapping values with half of its neighbors pointed by the ``red" arrows ($1 \sim 9$ directions for D3Q19 in Fig.\ref{fig:3D-swap-stream}),
if those neighbors are already in post-collision and have ``reverted" their distributions. 
We define this operation as ``$swap\_stream$". 
The ``\textbf{$revert$}" operation in Fig.\ref{fig:3D-collide-revert} lets a cell locally swap its post-collision distributions to opposite directions.
To make the Fuse Swap LBM more efficient, \added[]
{Palabos pre-processes and post-processes the boundary cells on the bounding box at line 2 and 7, respectively, so that it can remove the boundary checking operation in the inner bulk domain.
Thus Alg.\ref{alg:3D-fuse} is divided into three stages in every time step as follows.}


\begin{algorithm}[h!]
	\caption{3D Fuse Swap LBM}
	\label{alg:3D-fuse}
	\scriptsize{
	\begin{algorithmic}	[1]
		\For {iT = 0; iT $<$ N; ++iT}
		\State Stage I: $collide$ and $revert$ on the bounding box, i.e., 6 surfaces of cuboid (1,1,1) to ($lx,ly,lz$)
		\Statex // Stage II: bulk domain computation
		\For {iX = 2; iX $\leq$ $lx-1$; ++iX}
			\For {iY = 2; iY $\leq$ $ly-1$; ++iY}
			\For {iZ = 2; iZ $\leq$ $lz-1$; ++iZ}
				\State $collide$ \& $swap\_stream$ on (iX, iY, iZ) to half of its neighbors
			\EndFor
		\EndFor
		\EndFor
		\State Stage III: $boundary\_swap\_stream$ on the bounding box
		\EndFor
	\end{algorithmic}
	}
\end{algorithm}

\section{The 3D Memory-aware LBM Algorithm}
\label{sec:3D-mem-aware}



\subsection{Sequential 3D Memory-aware LBM}
\label{sec:3D-seq-mem-aware}

\begin{figure}[h!]
\captionsetup{justification=raggedright,format=hang}
     \centering
     \begin{subfigure}[t]{0.16\textwidth}
         \centering
         \includegraphics[width=\textwidth]{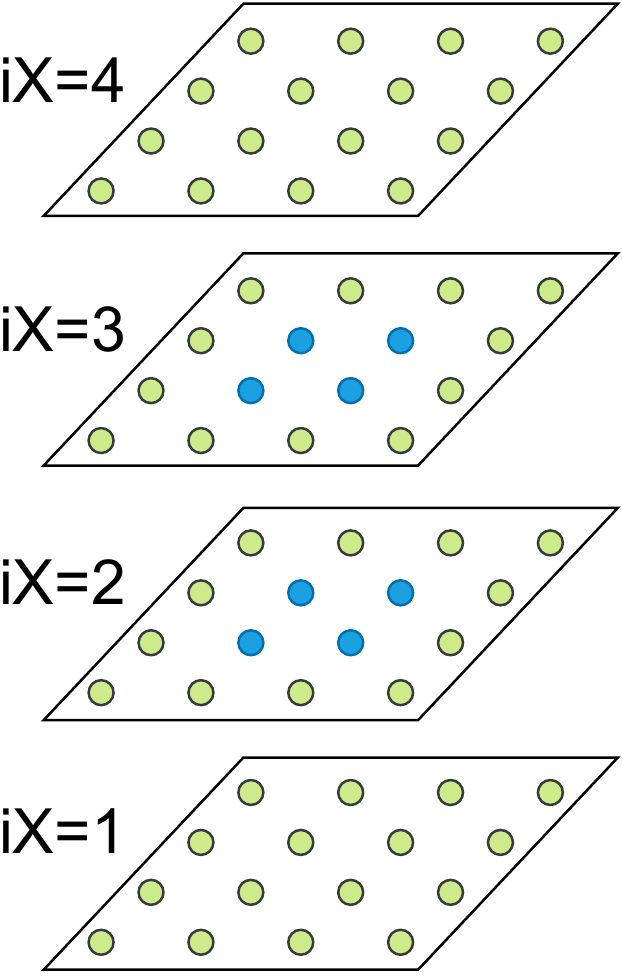}
         \caption{\scriptsize Initialization.}
         \label{fig:3D-2step-1}
     \end{subfigure}
     \begin{subfigure}[t]{0.16\textwidth}
         \centering
         \includegraphics[width=\textwidth]{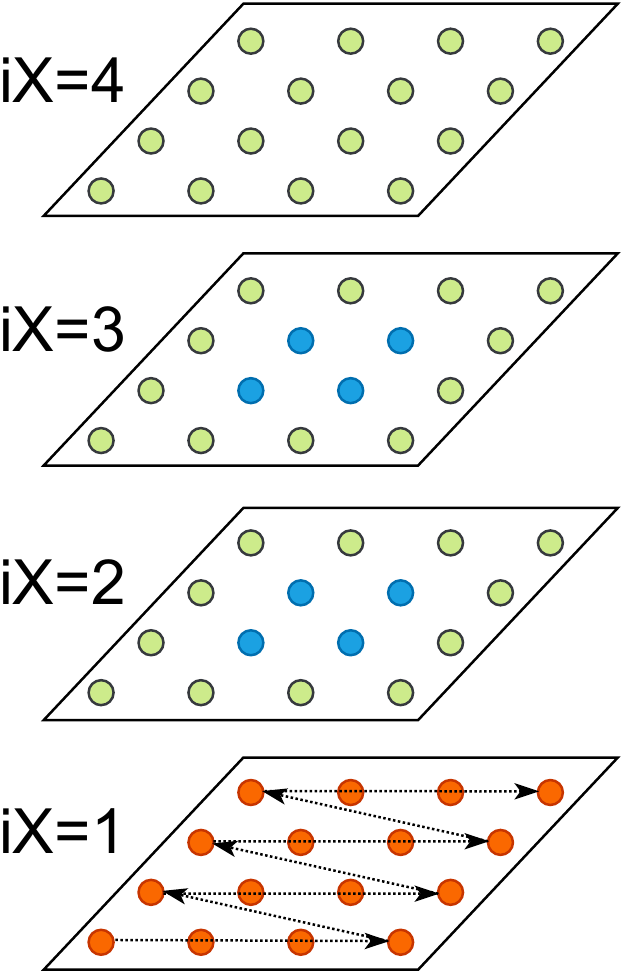}
         \caption{\scriptsize First computation on layer iX=1.} 
         \label{fig:3D-2step-2}
     \end{subfigure}
     \begin{subfigure}[t]{0.16\textwidth}
         \centering
         \includegraphics[width=\textwidth]{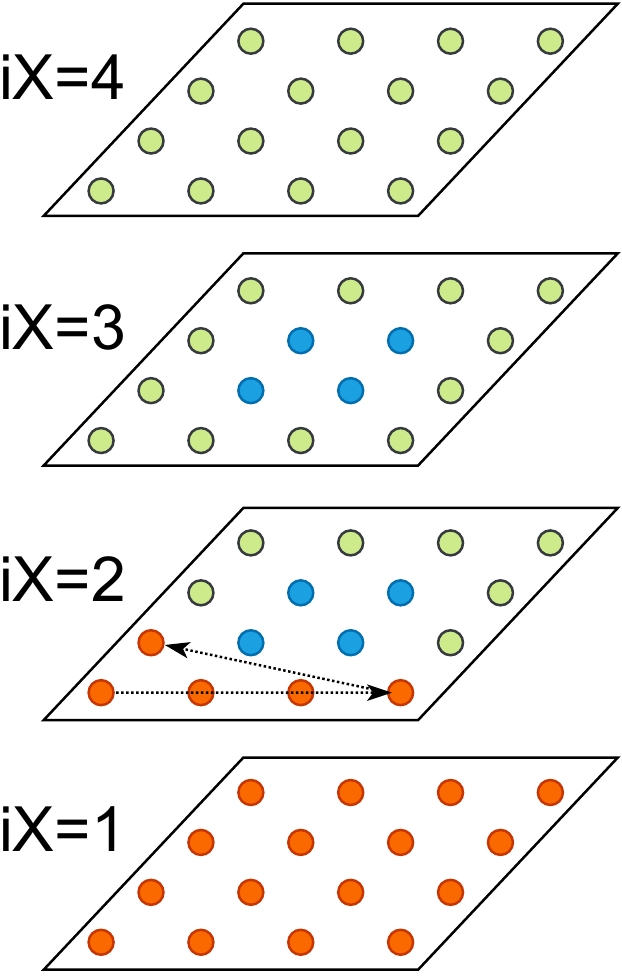}
         \caption{\scriptsize First computation on layer iX=2.} 
         \label{fig:3D-2step-3}
     \end{subfigure}
     \begin{subfigure}[t]{0.16\textwidth}
         \centering
         \includegraphics[width=\textwidth]{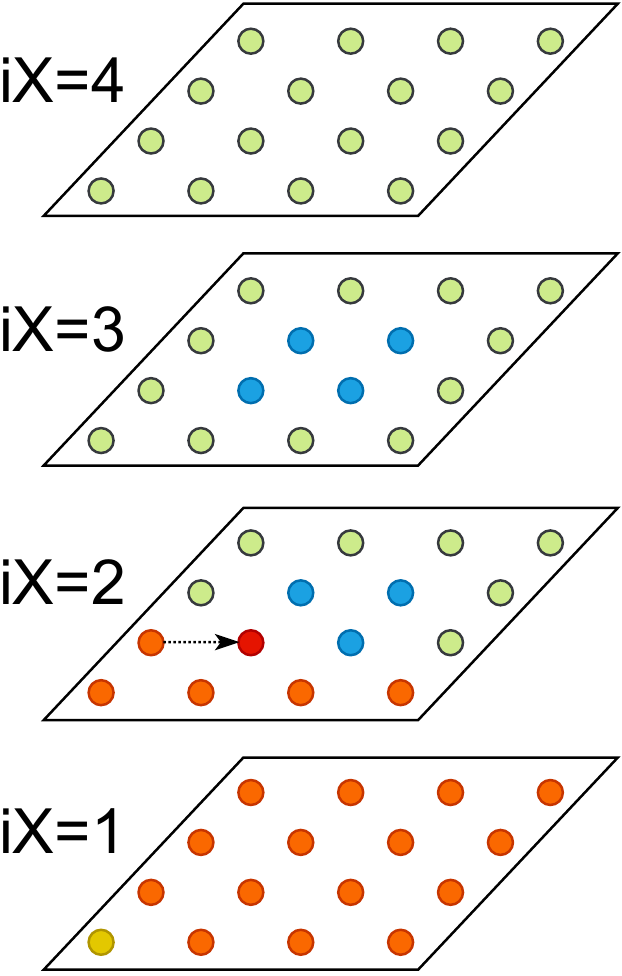}
         \caption{\scriptsize First computation on cell (2,2,2).} 
         \label{fig:3D-2step-4}
     \end{subfigure}
     \begin{subfigure}[t]{0.16\textwidth}
         \centering
         \includegraphics[width=\textwidth]{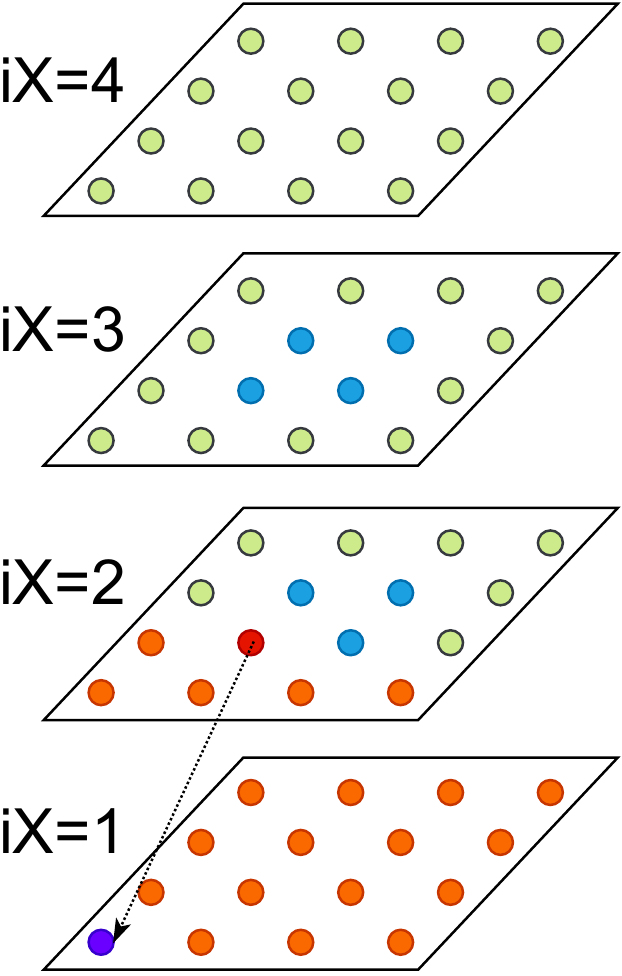}
         \caption{\scriptsize Second computation on cell (1,1,1).} 
         \label{fig:3D-2step-5}
     \end{subfigure}
     \begin{subfigure}[t]{0.16\textwidth}
         \centering
         \includegraphics[width=\textwidth]{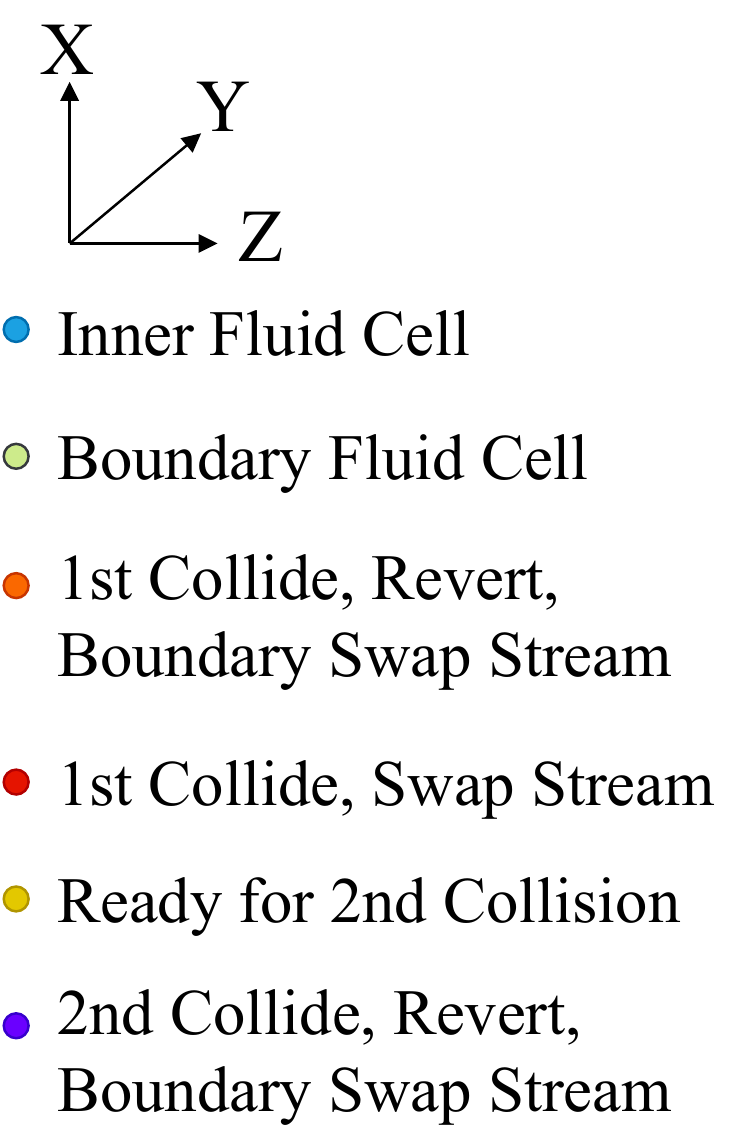}
         \caption{\scriptsize Legends.}
         \label{fig:3D-2step-6}
     \end{subfigure}
    \caption{\small 3D sequential two-step memory-aware LBM on a $4\times4\times4$ cube lattice.}
    \label{fig:3D-2step-seq}
\end{figure}

We design and develop the sequential 3D memory-aware LBM (shown in Alg.\ref{alg:3D-2step-seq-prism}),
based on the latest efficient Fuse Swap LBM, 
by adding two more features: merging two collision-streaming cycles to explore the temporal locality, 
and introducing the prism traversal to explore the spatial locality.
%
Fig.\ref{fig:3D-2step-seq} shows an example on
how to 
merge two collision-streaming cycles given a $4\times4\times4$ cube:
\begin{enumerate}
\itemsep0em

\item Fig.\ref{fig:3D-2step-1} shows the initial state of all cells at the current time step $t$.
Green cells are on boundaries, and blue cells are located in the inner bulk domain.

\item In Fig.\ref{fig:3D-2step-2}, we compute the first $collide$, $revert$ and $boundary\_swap\_stream$ row by row 
on the bottom layer iX = 1. 
After a cell completes the first computation, we change it to orange.

\item In Fig.\ref{fig:3D-2step-3}, we compute the first $collide$ and $boundary\_swap\_stream$ row by row till cell (2,2,1) on the second layer iX = 2.

\item In Fig.\ref{fig:3D-2step-4}, cell (2,2,2) completes its first $collide$ and $swap\_stream$, so we change it to red since they are inner cells. Then we observe that cell (1,1,1) is ready for the second $collide$, so we change it to yellow.

\item In Fig.\ref{fig:3D-2step-5}, we execute the second $collide$ and $boundary\_swap\_stream$ on cell (1,1,1), and change it to purple.
\end{enumerate}

\begin{figure}[b!]
     \centering
     \begin{subfigure}[t]{0.24\textwidth}
         \centering
         \includegraphics[width=\textwidth]{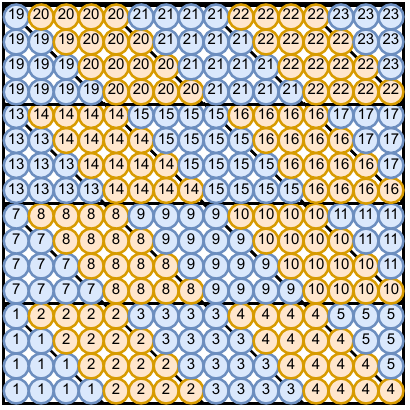}
         \caption{\small Layer iX=1.}
         \label{fig:3D-seq-prism-1}
     \end{subfigure}
     \begin{subfigure}[t]{0.24\textwidth}
         \centering
         \includegraphics[width=\textwidth]{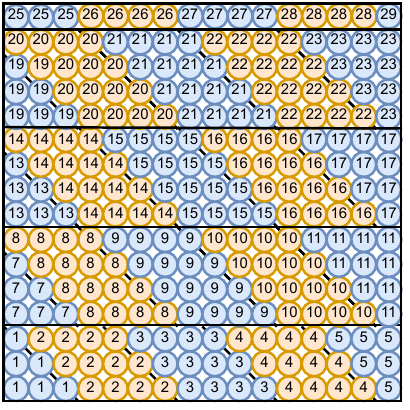}
         \caption{\small Layer iX=2.}
         \label{fig:3D-seq-prism-2}
     \end{subfigure}
     \begin{subfigure}[t]{0.24\textwidth}
         \centering
         \includegraphics[width=\textwidth]{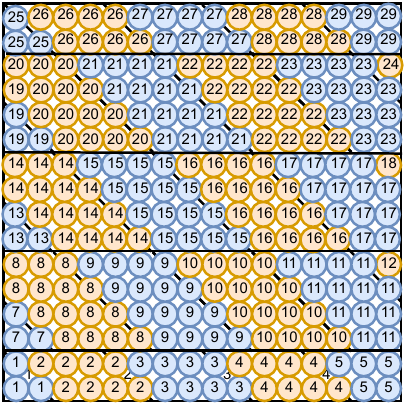}
         \caption{\small Layer iX=3.}
         \label{fig:3D-seq-prism-3}
     \end{subfigure}
     \begin{subfigure}[t]{0.24\textwidth}
         \centering
         \includegraphics[width=\textwidth]{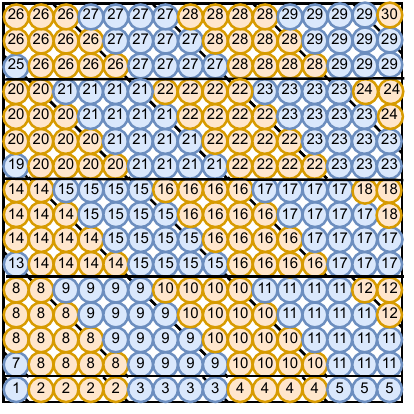}
         \caption{\small Layer iX=4.}
         \label{fig:3D-seq-prism-4}
     \end{subfigure}
     \begin{subfigure}[t]{0.48\textwidth}
         \centering
         \includegraphics[width=0.5\textwidth]{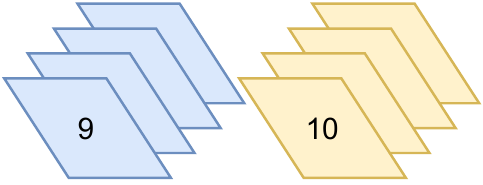}
         \caption{\small Prism 9 and 10 are parallelpiped shape. Layer iX=4 is on the top.}
         \label{fig:3D-seq-prism-9-10}
     \end{subfigure}
     \quad
     \centering
     \begin{subfigure}[t]{0.48\textwidth}
         \centering
         \includegraphics[width=0.4\textwidth]{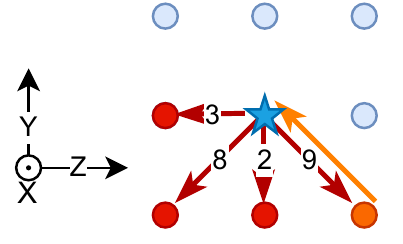}
         \caption{Planar slice when cutting Fig.\ref{fig:3D-swap-stream} (swap stream operation) along Y-Z plane.}
	    \label{fig:3D-swap-stream-yz}
     \end{subfigure}
    \caption{Sequential 3D prism traversal on a $4 \times 16 \times 16$ cuboid box.}
    \label{fig:3D-seq-prism}
\end{figure}

To further increase data reuse, we optimize the algorithm's spatial locality by designing a ``prism traversal" method, since the shape of this traversal constructs a 3D pyramid prism or a parallelpiped prism. 
We use an example to explain its access pattern in a $4 \times 16 \times 16$ cuboid with stride $tile=4$.
Fig.\ref{fig:3D-seq-prism-1}$\sim$\ref{fig:3D-seq-prism-4} are the four separate $16 \times 16$ layers of the cuboid from bottom to top.
The cells with the same number on the four layers construct a \textit{prism} (e.g., the cells with number 1 in Fig.\ref{fig:3D-seq-prism-1}$\sim$\ref{fig:3D-seq-prism-4} construct a pyramid-shape ``Prism 1").
In each prism, we still firstly go along Z-axis, then along Y-axis, and upward along X-axis at last.
Then we traverse prism-wise from Prism 1 to Prism 30.
Finally, if a cuboid is much larger than this example, 
the majority of prisms are ``parallelpiped" shapes like Prism 9 and 10 in Fig.\ref{fig:3D-seq-prism-9-10}. 
The reason why the planar slice of a prism is either triangles or parallelograms is due to the $swap\_stream$ operation.
When cutting Fig.\ref{fig:3D-swap-stream} ($swap\_stream$) along the Y-Z plane, we have a planar slice as shown in Fig.\ref{fig:3D-swap-stream-yz}.
We observe that a cell (star) swaps with its lower right neighbor (orange) at direction 9.
In other words, when the orange cell swaps with the upward row, 
its neighbor ``shifts" one cell \textit{leftward}.
Similarly, if cutting Fig.\ref{fig:3D-swap-stream} ($swap\_stream$) along the X-Y plane,
when a cell swaps data with the upward row, its neighbor ``shifts" one cell \textit{forward}. 
Thus when we traverse $tile$ number of cells on Z-axis at row $iY$, 
they can swap with $tile$ number of cells but shifted one cell leftward at row $iY+1$, 
thereby we get parallelograms in Fig.\ref{fig:3D-seq-prism-1}$\sim$\ref{fig:3D-seq-prism-4}. 
When the shift encounters domain boundaries, we truncate the parallelograms and get isosceles right triangles or part of parallelograms. At last, we can safely combine ``prism traversal" with merging two collision-streaming cycles, since the cell at left forward down corner has been in a post-collision state and ready to compute the second computation when following the above traversal order.

\begin{algorithm}[h!]
	\caption{3D Sequential Memory-aware LBM}\label{alg:3D-2step-seq-prism}
	\scriptsize{
	\begin{algorithmic}	[1]
	\State tile := stride of the prism traversal
	\For {iT = 0; iT $<$ N; iT += 2}
		
		\For {outerX = 1; outerX $\leq$ $lx$; outerX += tile}
        \For {outerY = 1; outerY $\leq$ $ly$ + tile - 1; outerY += tile}
        \For {outerZ = 1; outerZ $\leq$ $lz$ + 2* (tile - 1); outerZ += tile}
        \For {innerX=outerX; innerX $\leq$ MIN(outerX+tile-1, $lx$); ++innerX, ++dx}
        \State minY = outerY - dx; maxY = minY + tile - 1; dy = 0; \added[]{/* forward shift */}
        \For {innerY=MAX(minY, 1); innerY $\leq$ MIN(maxY, $ly$); ++innerY, ++dy}
        \State minZ = outerZ - dx - dy; maxZ = minZ + tile - 1; \added[]{/* leftward shift */}
        \For {innerZ=MAX(minZ, 1); innerZ $\leq$ MIN(maxZ, $lz$); ++innerZ}
            \Statex \hspace*{1cm}/* (1) First computation at time step $t$. */
            \State $adaptive\_collide\_stream$(innerX, innerY, innerZ);
			
			\Statex \hspace*{1cm}/* (2) Second computation at time step $t+1$. */
			\If{innerX $>$ 1 \&\& innerY $>$ 1 \&\& innerZ $>$ 1}
                \State $adaptive\_collide\_stream$(innerX-1, innerY-1, innerZ-1);
			\EndIf
			
			\Statex \hspace*{1cm}/* (3) Second computation of neighbors at certain locations. */
			\State $boundary\_neighbor\_handler$(innerX, innerY, innerZ);
			
			
    			
		\EndFor
		\EndFor
		\EndFor
		\EndFor
		\EndFor
		\EndFor
	\State Second $collide$, $revert$ \& \textit{boundary\_swap\_stream} on the top layer iX = $lx$.
	\EndFor
	
	\Function {$boundary\_cell\_comp$}{iX, iY, iZ}
	\State $collide$, $revert$, \& \textit{boundary\_swap\_stream} on (iX, iY, iZ) to half of its neighbors;
	\EndFunction
	
	\Function {$adaptive\_collide\_stream$}{iX, iY, iZ}
	\If{(iX, iY, iZ) is on the boundary}
        \State $boundary\_cell\_comp$(iX, iY, iZ);
    \Else
    	    \State $collide$ \& $swap\_stream$ on (iX, iY, iZ) to half of its neighbors;
    \EndIf
	\EndFunction
	
	\Function {$boundary\_neighbor\_handler$}{iX, iY, iZ}
	\Statex // Handle the second computation of (iX, iY, iZ)'s neighbors at certain locations.
	\If{iZ $==$ $lz$} // (iX, iY, iZ) is the last cell of a row.
    	\State $boundary\_cell\_comp$ (iX-1, iY-1, iZ);
    \EndIf
    
    \If{iY $==$ $ly$ \&\& iZ $>$ 1} // (iX, iY, iZ) is in the last row of a layer.			
    	\State $boundary\_cell\_comp$(iX-1, iY, iZ-1);
    \EndIf
    
    \If{iY $==$ $ly$ \&\& iZ $==$ $lz$} // (iX, iY, iZ) is the last cell on a layer.			
    	\State $boundary\_cell\_comp$(iX-1, iY, iZ);
    \EndIf
	\EndFunction
	\end{algorithmic}
	}
\end{algorithm}
Alg.\ref{alg:3D-2step-seq-prism} presents the sequential 3D memory-aware LBM.
Lines $6\sim10$ traverse the domain prism-wise with stride $tile$.
Lines $11\sim14$ merge two time steps computation.
The first $stream$ starting from the bottom layer iX = 1 in Line 11 is necessary due to the data dependency for the second computation.
In particular, the if-statement in Line 13 ensures that the cell to compute at time step $t+1$ is in a post-collision state, no matter using D3Q15, D3Q19, D3Q27 or extended lattice models.
For simplicity, Lines 16$\sim$29 define three helper functions. 

	
	
    
    

\subsection{Parallel 3D Memory-aware LBM}
\label{sec:3D-omp-mem-aware-prism}
\begin{figure}[h!]
     \centering
     \begin{subfigure}[t]{0.205\textwidth}
         \centering
         \includegraphics[width=\textwidth]{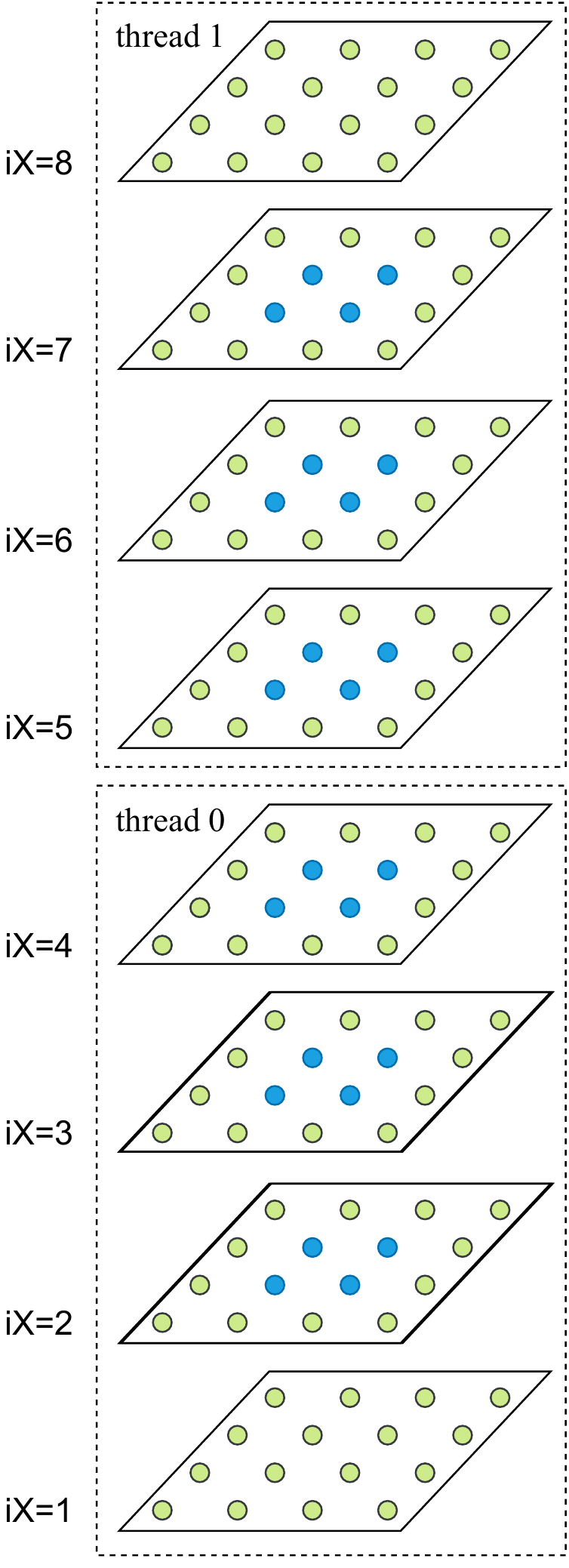}
         \caption{\small Initialization.}
         \label{fig:3D-2step-prism-omp-1}
     \end{subfigure}
     \begin{subfigure}[t]{0.235\textwidth}
         \centering
         \includegraphics[width=\textwidth]{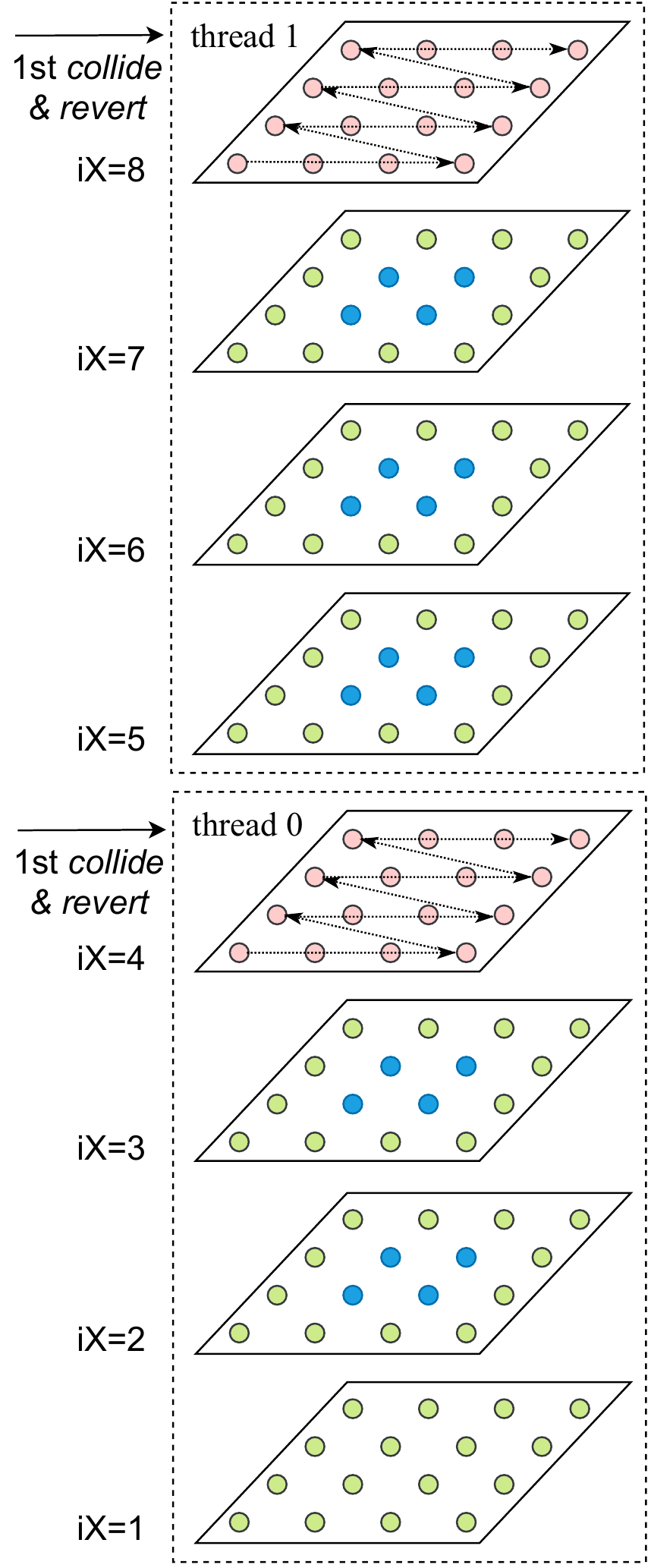}
         \caption{\small Stage I.} 
         \label{fig:3D-2step-prism-omp-2}
     \end{subfigure}
     \begin{subfigure}[t]{0.24\textwidth}
         \centering
         \includegraphics[width=\textwidth]{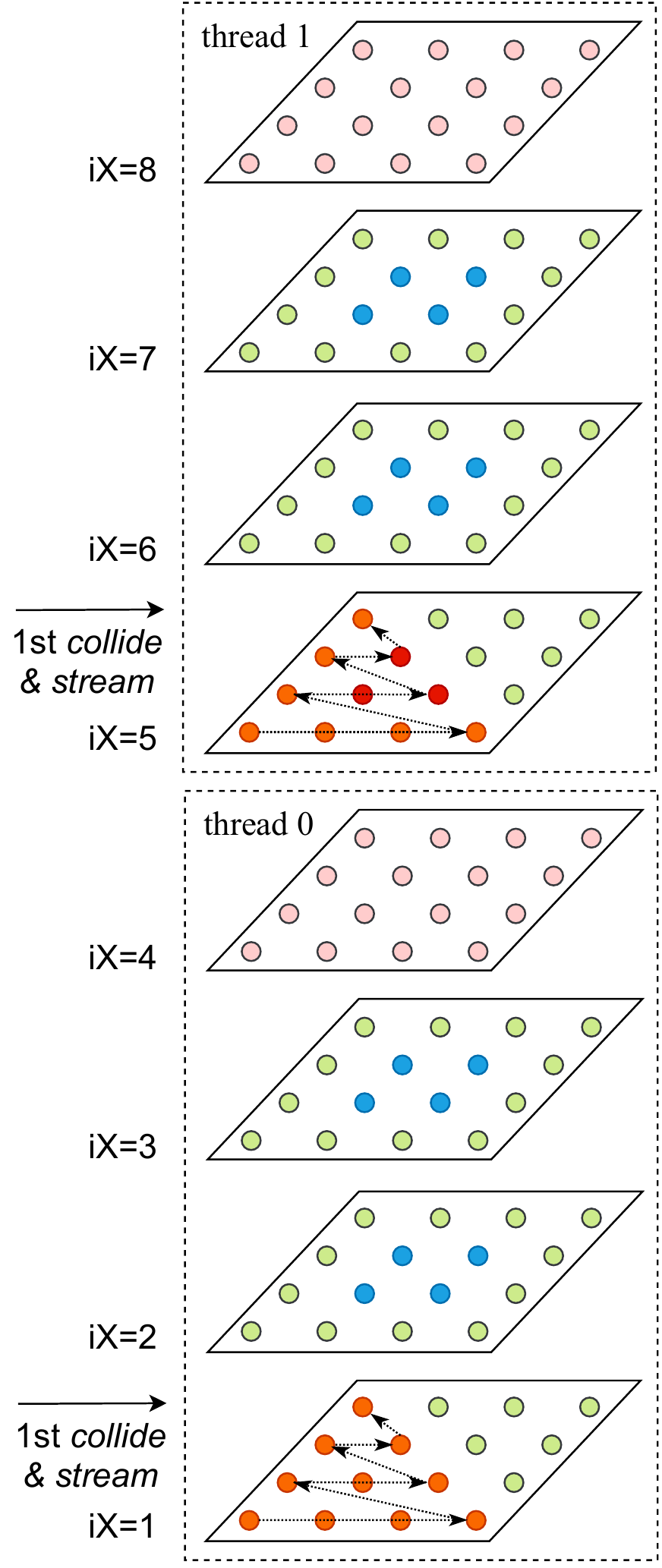}
         \caption{\small Stage II: Case 1.} 
         \label{fig:3D-2step-prism-omp-3}
     \end{subfigure}
     \begin{subfigure}[t]{0.24\textwidth}
         \centering
         \includegraphics[width=\textwidth]{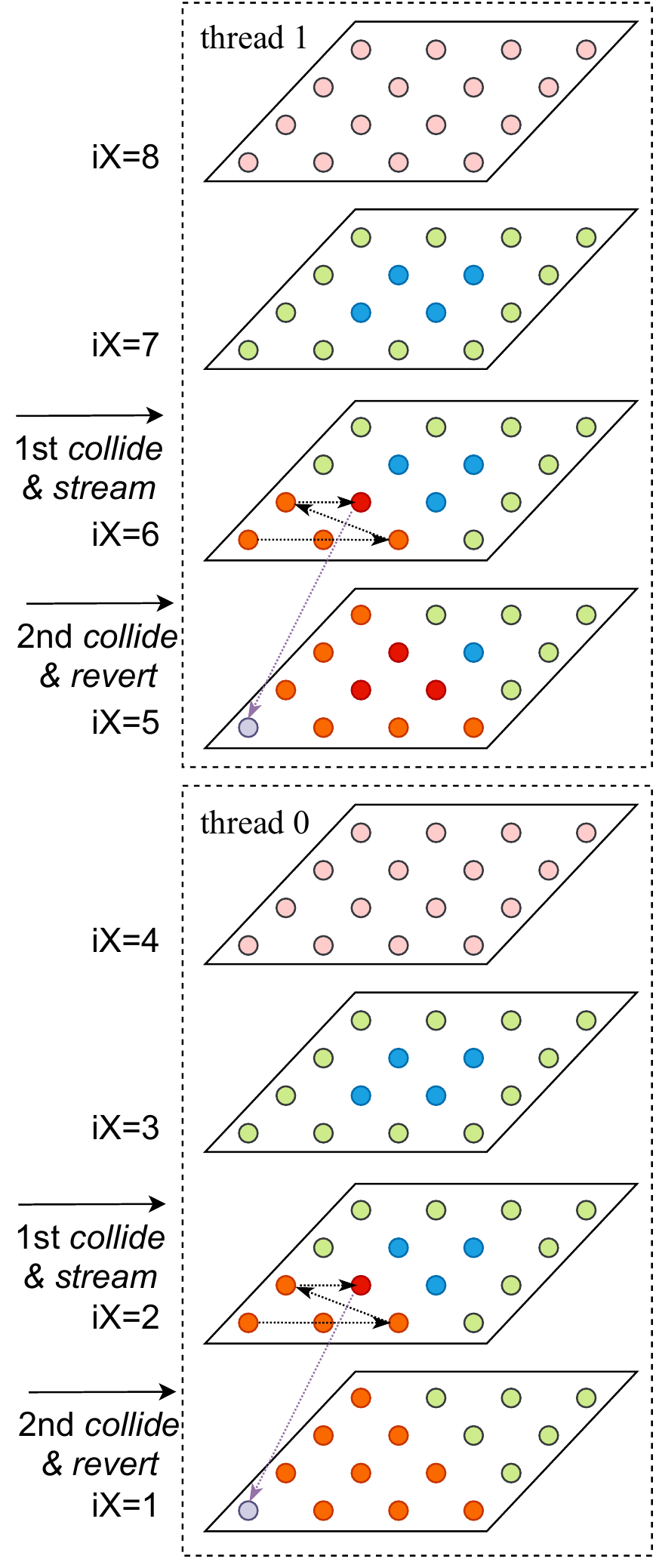}
         \caption{\small Stage II: Case 2.} 
         \label{fig:3D-2step-prism-omp-case-2}
     \end{subfigure}
       \begin{subfigure}[t]{0.24\textwidth}
         \centering
         \includegraphics[width=\textwidth]{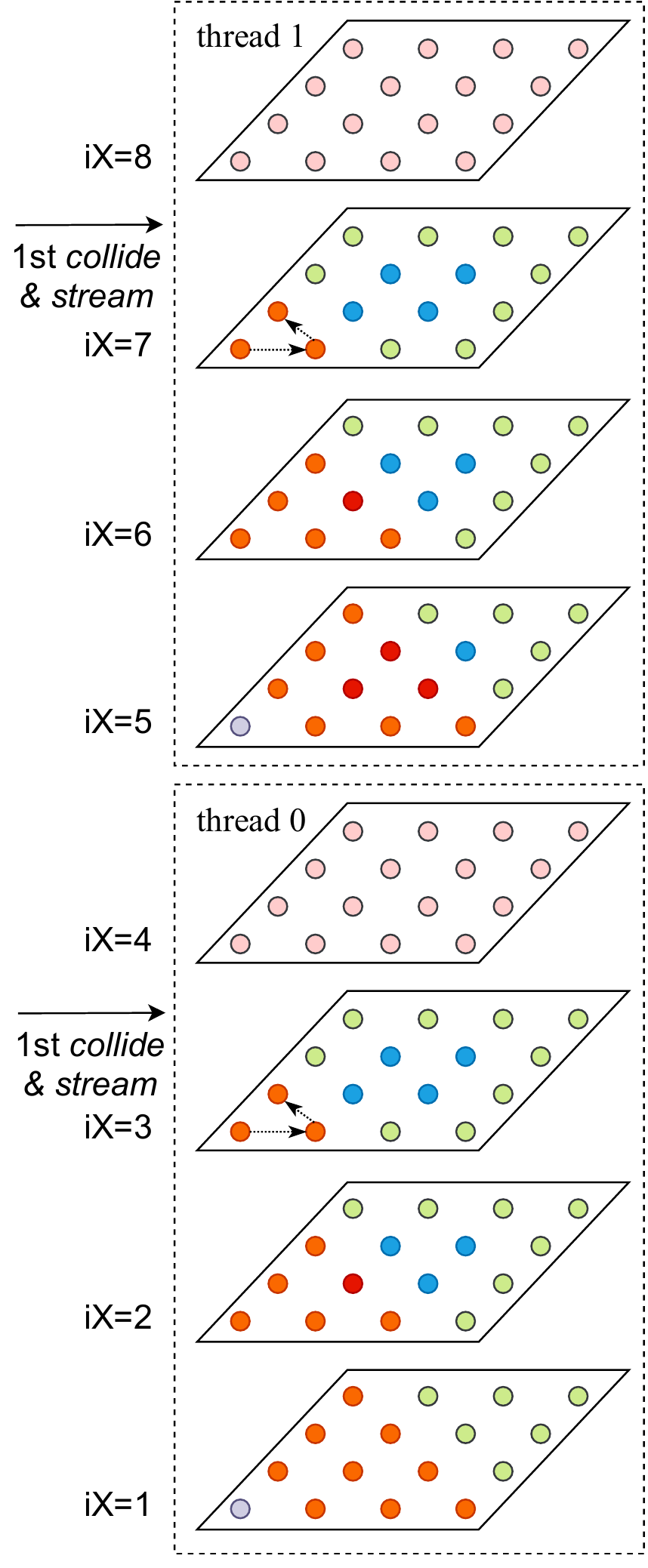}
         \caption{\small Stage II: Case 0.}
         \label{fig:3D-2step-prism-omp-case-0}
     \end{subfigure}
     \begin{subfigure}[t]{0.24\textwidth}
         \centering
         \includegraphics[width=\textwidth]{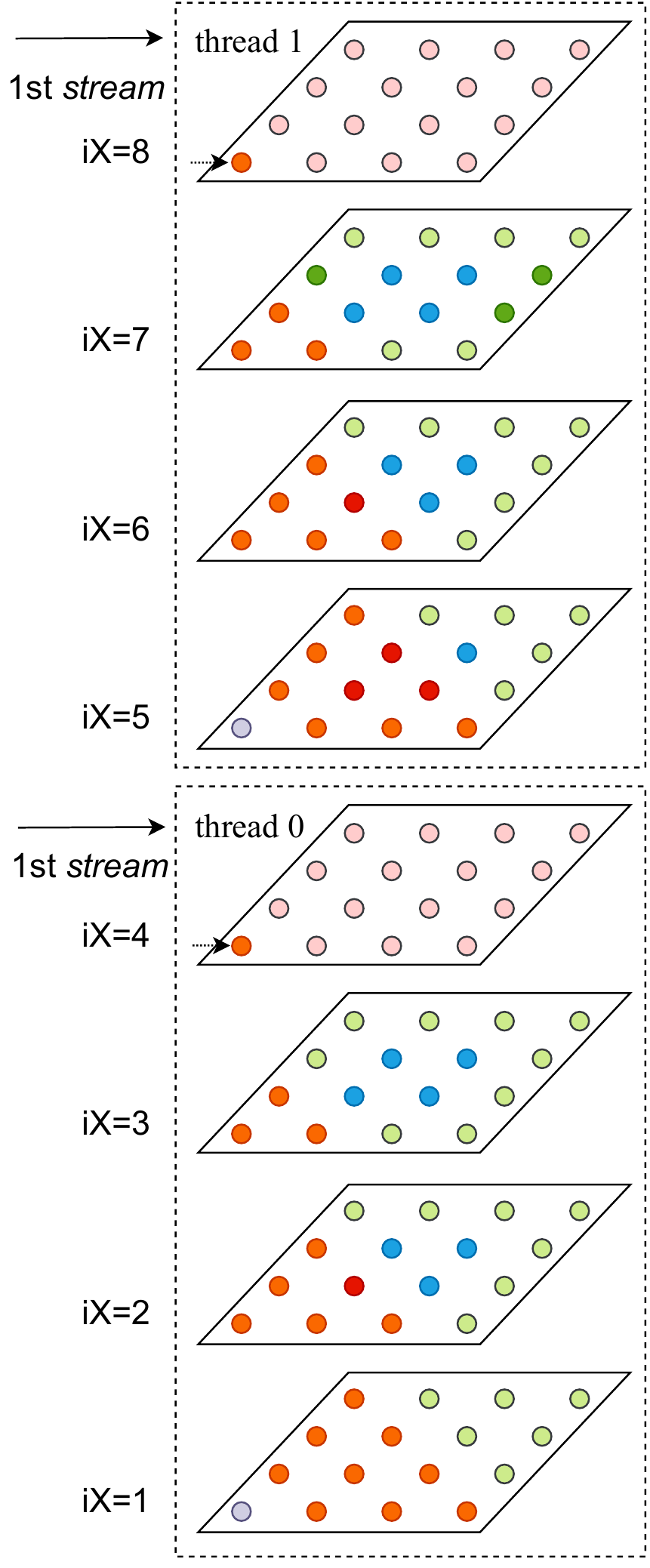}
         \caption{\small Stage II: Case 3.} 
         \label{fig:3D-2step-prism-omp-7}
     \end{subfigure}
     \begin{subfigure}[t]{0.24\textwidth}
         \centering
         \includegraphics[width=\textwidth]{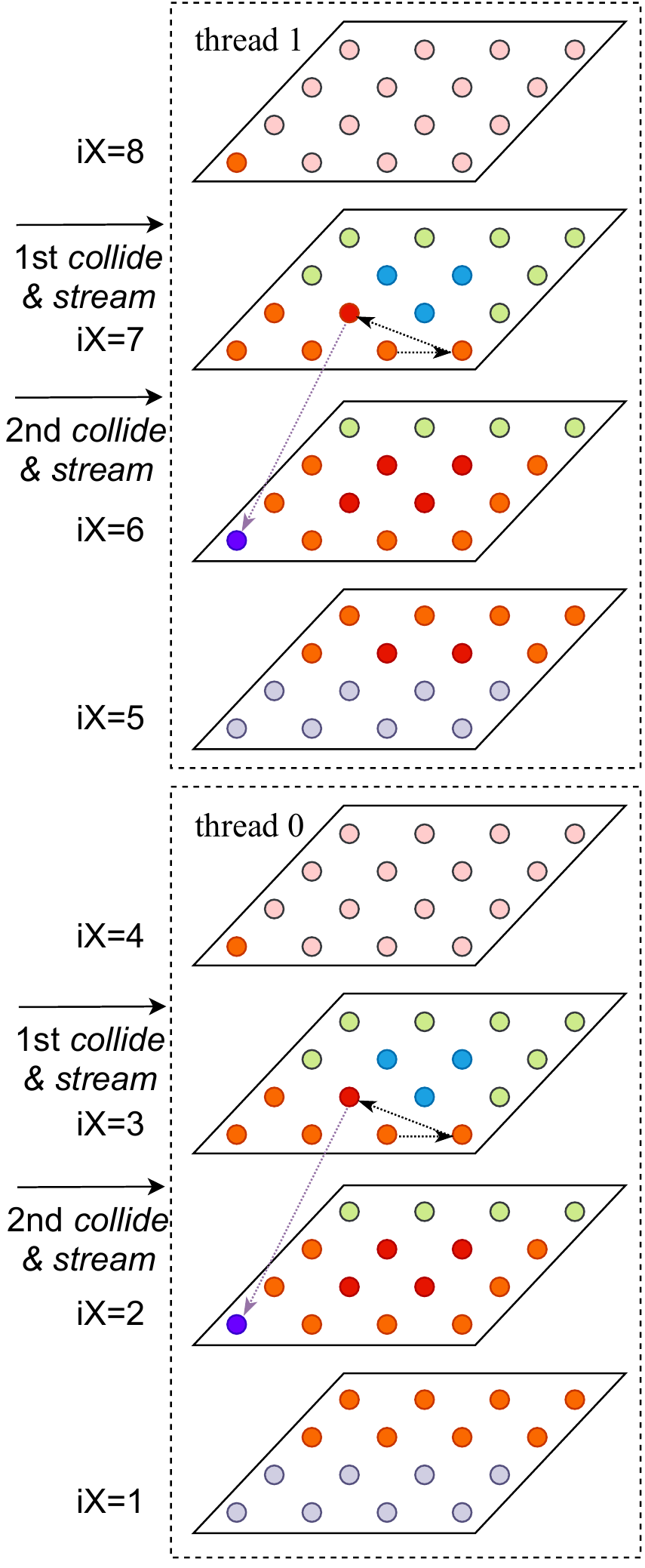}
         \caption{\small Stage II: Case 4} 
         \label{fig:3D-2step-prism-omp-12}
     \end{subfigure}
     \begin{subfigure}[t]{0.24\textwidth}
         \centering
         \includegraphics[width=\textwidth]{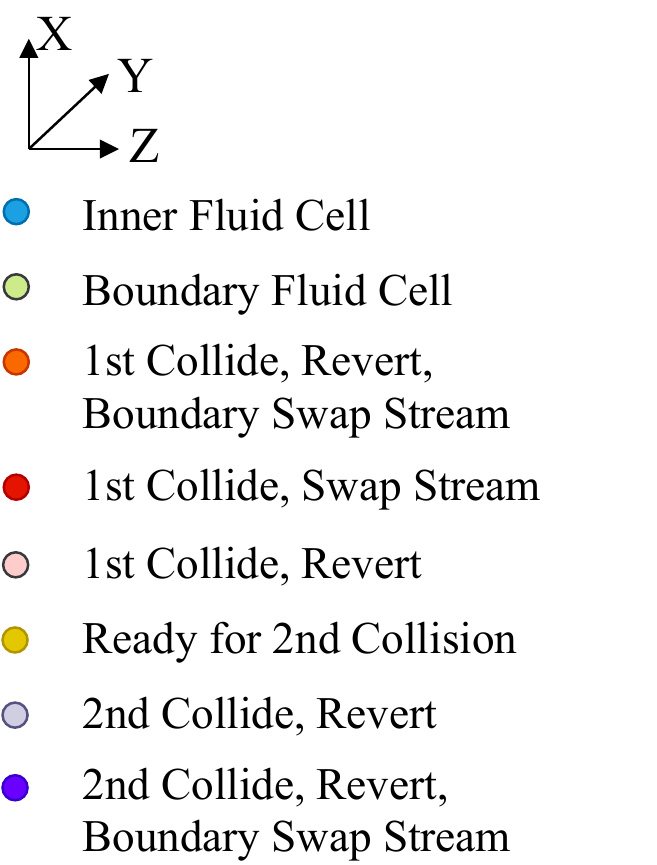}
         \caption{\small Legends.}
         \label{fig:3D-2step-prism-omp-legend}
     \end{subfigure}
    \caption{\small Parallel 3D two-step memory-aware LBM on a $8\times4\times4$ cuboid.}
    \label{fig:3D-2step-prism-omp-A}
\end{figure}

To support manycore systems, we choose OpenMP~\cite{OPENMP} to realize the parallel 3D memory-aware LBM algorithm~\footnote{\cite{slaughter2020task} states that when the minimum effective task granularity (METG) of parallel runtime systems are smaller than task granularity of large-scale LBM simulations, all of these runtime system can deliver good parallel performance.}.
%
Fig.\ref{fig:3D-2step-prism-omp-A} illustrates its idea on a $8\times4\times4$ cuboid, which is evenly partitioned by two threads along the X-axis (\textit{height}).
Then each thread traverses a $4\times4\times4$ sub-domain with prism stride $tile=4$.
Line 4 in Alg.\ref{alg:3D-2step-prism-omp} defines the start and end layer index of each thread's sub-domain, thus the end layers $myEndX$ are ``\textit{intersections}" (e.g., layer 4 and 8).
Fig.\ref{fig:3D-2step-prism-omp-1} shows the initial state at time step $t$.
In addition, the parallel 3D memory-aware Alg.\ref{alg:3D-2step-prism-omp} \added[]{consists of} three stages: Preprocessing, Sub-domain computation, and Post-processing.



\begin{algorithm}[h!]
	\caption{ Parallel 3D Memory-aware LBM}
	\label{alg:3D-2step-prism-omp}
	\scriptsize{
	\begin{algorithmic}	[1]
	
	\For {iT = 0; iT $<$ N; iT += 2}
	
	\State \textbf{\#pragma omp parallel default(shared)\{}
	\State $sub\_h = lx / nthreads$; // height of each thread's sub-domain
	\State myStartX = $1 + thread\_id \times sub\_h$; myEndX = $(thread\_id+1) \times sub\_h$; 
	\Statex \textit{/* Stage I: First collide \& revert on the intersection layer.*/}
    \State $collide$ \& $revert$ on all $ly\times lz$ cells on layer iX = myEndX;
	\State \textbf{\#pragma omp barrier}

	\Statex \textit{/* Stage II: Main computation in each thread's sub-domain.*/}
	\For {outerX = myStartX; outerX $\leq$ myEndX; outerX += tile}
    \For {outerY = 1; outerY $\leq$ $ly$ + tile - 1 ; outerY += tile}
    \For {outerZ = 1; outerZ $\leq$ $lz$ + 2 * (tile - 1); outerZ += tile}
    
    \For {innerX=outerX; innerX$\leq$MIN(outerX+tile-1, myEndX); ++innerX, ++dx}
    \State minY = outerY - dx; maxY = minY + tile - 1; dy = 0; \added[]{/* forward shift */}
    \For {innerY=MAX(minY, 1); innerY$\leq$MIN(maxY, $ly$); ++innerY, ++dy}
    \State minZ = outerZ - dx - dy; maxZ = minZ + tile - 1; \added[]{/* leftward shift */}
    \For {innerZ = MAX(minZ, 1); innerZ $\leq$ MIN(maxZ, $lz$); ++innerZ} 
    
        \Statex // \textit{Case 0: First $collide$ \& $stream$ on the first row and column of each layer except the intersection layers.}
        \If{innerX != myEndX \&\& (innerX == 1 or innerY == 1 or innerZ == 1)}
        \State First $boundary\_cell\_comp(innerX, innerY, innerZ)$; 
        \State \textbf{continue;}
        \EndIf

        \Statex // \textit{Case 1: First $collide$ \& $stream$ on layer $myStartX$:}
        \If{innerX == myStartX}
        \State First $adaptive\_collide\_stream$(innerX, innerY, innerZ);
  
        \Statex // \textit{Case 2: First $collide$ \& $stream$ on $myStartX + 1$; Second $collide$ \& $revert$ on $myStartX$:}
        \ElsIf{innerX == myStartX + 1}
            \State First $adaptive\_collide\_stream$(innerX, innerY, innerZ);
			
			\State Second $collide$ \& $revert$ on (innerX-1, innerY-1, innerZ-1);
			\State Handle the second $collide$ \& $revert$ of neighbors at certain boundary locations;
        
        \Statex // \textit{Case 3: First $stream$ on layer $myEndX$; Second $collide$ \& $stream$ under one layer:}
        \ElsIf{innerX == myEndX}
            \State First $adaptive\_stream$(innerX, innerY, innerZ);
			\State Second $adaptive\_collide\_stream$(innerX-1, innerY-1, innerZ-1);
			\State $boundary\_neighbor\_handler$ (innerX, innerY, innerZ);
        
        \Statex // \textit{Case 4: first $collide$ \& $stream$ on other layers; Second $collide$ \& $stream$ under one layer:}
		\Else

            \State First $adaptive\_collide\_stream$(innerX, innerY, innerZ);
			\State Second $adaptive\_collide\_stream$(innerX-1, innerY-1, innerZ-1);
			
			\State  $boundary\_neighbor\_handler$(innerX, innerY, innerZ);
        \EndIf
	\EndFor
	\EndFor
	\EndFor
	\EndFor
	\EndFor
	\EndFor
    
    \State \textbf{\#pragma omp barrier}
    
    \Statex \textit{/* Stage III: second $collide$ \& $stream$ on the intersection; then second $stream$ on the  layer $myStartX$. */}
    \State $adaptive\_collide\_stream$ at all $ly\times lz$ cells on layer iX = myEndX;
	\State \textbf{\#pragma omp barrier}
	
    \State $stream$ at all $ly\times lz$ cells on layer iX = myStartX;
	
	\State	\textbf{\}}
	\EndFor

	\end{algorithmic}
	}
\end{algorithm}

\begin{enumerate}
\itemsep0em
	\item \textbf{Stage I (Preprocessing)} \textit{line 5 in Alg.\ref{alg:3D-2step-prism-omp}}: 
	In Fig.\ref{fig:3D-2step-prism-omp-2}, thread 0 and 1 compute the first $collide$ and $revert$ on the ``intersection" layers 4 and 8, respectively, 
	and then change them to pink.
	
	\item \textbf{Stage II (Sub-domain computation)} handles five cases from step 2 to 7. 
	In \textit{case 0} (\textit{lines 15$\sim$17 in Alg.\ref{alg:3D-2step-prism-omp}}), when thread 0 and 1 access the cells on the first row and column of each layer except the ``intersection" layers, 
	we execute the first $boundary\_cell\_comp$ on them and change them to orange.
	
	\item Fig.\ref{fig:3D-2step-prism-omp-3} shows \textit{case 1} (\textit{lines 18$\sim$19 in Alg.\ref{alg:3D-2step-prism-omp}}). 
	When thread 0 and 1 access the cells on layer $myStartX$ (iX = 1 \& 5), respectively, 
	we execute the $adaptive\_collide\_stream$ on them to compute at time step $t$,
	and then change the boundary cells to orange and the inner cells to red.

	\item Fig.\ref{fig:3D-2step-prism-omp-case-2} shows \textit{case 2} (\textit{lines 20$\sim$23 in Alg.\ref{alg:3D-2step-prism-omp}}). 
	When thread 0 and 1 are on layer $myStartX+1$ (iX = 2 \& 6), respectively, 
	we execute the first $adaptive\_collide\_stream$ at time step $t$
	and change boundary cells to orange and inner cells to red.
	Meanwhile, cell (5,1,1) and (1,1,1) 
	have collected the data dependencies to $collide$ at time step $t+1$, 
	we execute the second $collide$ and $revert$ but without $stream$ on them, and change to light purple.

    \item Fig.\ref{fig:3D-2step-prism-omp-case-0} shows that when continuing traversal in Prism 1,
    thread 0 and 1 are on layer iX = 3 \& 6. 
    Since the cells traversed in this figure are in the first row and column, case 0 is used here, otherwise, case 4 is used.

    \item Fig.\ref{fig:3D-2step-prism-omp-7} shows \textit{case 3} (\textit{lines 24$\sim$27 in Alg.\ref{alg:3D-2step-prism-omp}}). 
    When thread 0 and 1 are on the intersection layers (iX = 4 \& 8),
    we execute the remaining first $stream$ at time step $t$ due to preprocessing in Stage I.
    Then if cells under one layer (iX = 3 \& 7) collect their data dependency at time step $t+1$, 
    we execute the second $adaptive\_collide\_stream$ on them.
	
	
	\item Fig.\ref{fig:3D-2step-prism-omp-12} shows \textit{case 4} (\textit{lines 28$\sim$31 in Alg.\ref{alg:3D-2step-prism-omp}}). 
	When thread 0 and 1 are on the other layers of sub-domain, 
	we conduct the first \textit{adaptive\_collide\_stream} on (innerX, innerY, innerZ) at time step $t$, 
	and then the second \textit{adaptive\_collide\_stream} on (innerX-1, innerY-1, innerZ-1) at time step $t+1$.
	Then we call \textit{boundary\_neighbor\_handler} to compute the neighbors of (innerX, innerY, innerZ) at certain locations at time step $t+1$.
	
	
	\item \textbf{Stage III (Post-processing)} \textit{lines 33$\sim$35 in Alg.\ref{alg:3D-2step-prism-omp}}: 
	Firstly, since Stage I and case 3 have completed the first computation on intersection layers,
	we wrap up the second $collide$ and $stream$ on intersections.
	Secondly, since case 2 have executed the second $collide$ and $revert$ on the first layers $myStartX$ of each sub-domain, the second $stream$ remains to be executed.

\end{enumerate}


\textbf{How to Handle Thread Safety near Intersection Layers:}
\begin{figure}[t!]
	\centering
	\includegraphics[width=0.73\textwidth]{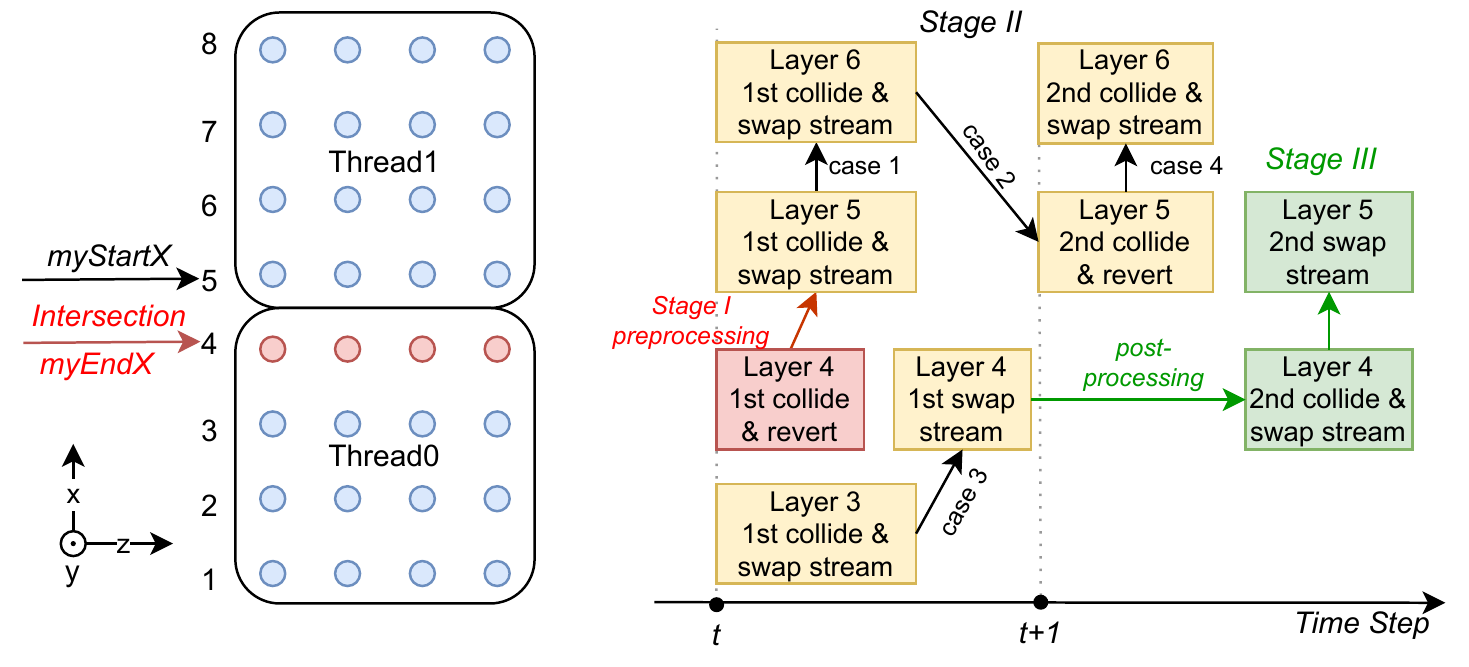}
	\caption{\small Handle thread safety on intersection layers. 
	}
	\label{fig:3D-omp-dependency}
\end{figure}
We aim to keep thread safety and minimize the synchronization cost \added[]{during parallel executions.
To this end,} we need to carefully design the initial state of each thread \added[comment={why max each thread's workload? We want to let each thread's sub-domain large enough, so threads will compute locally in most of time, then they synchronize less frequently only at the intersection layers.}]{so that the majority of computation stays in each threads' local sub-domain}.
The left part of Fig.\ref{fig:3D-omp-dependency} shows the view of Fig.\ref{fig:3D-2step-prism-omp-A} along X-Z axis, and layer 4 is the intersection layer that partitions two threads' sub-domains.
The right part shows the data dependencies near the intersection layer in two time steps.
In the figure, the red block represents Stage I of Alg.\ref{alg:3D-2step-prism-omp}, yellow blocks Stage II, and green blocks Stage~III .
The arrows indicate that data are transferred from layer A to B by using a procedure (or B depends on A).
\added[]{There are three non-trivial dependencies requiring to handle thread safety near intersection layers.}
(1) 
Since the swap algorithm only streams data to half of the neighbors under one layer,
the $swap\_stream$ on layer 5 ---the first layer of thread 1's sub-domain--- should be delayed after the $revert$ on layer 4 in thread 0's sub-domain. 
Thus, in Stage I, we pre-process $collide$ and $revert$ at time step $t$ but without $stream$ on layer 4, since 
$stream$ on layer 4 depends on the post-collision on layer 3, which has not been computed yet. 
(2) In Stage II, the second $swap\_stream$ on layer 6 called by the case 4 procedure should be delayed after the second $revert$ but without $swap\_stream$ on layer 5. 
This is because thread 1 cannot guarantee that thread 0 has completed the second $swap\_steam$ on layer 4. 
To keep thread safety, $swap\_stream$ on layer 5 is delayed to Stage III.
(3) Thus, in Stage III, the second $swap\_stream$ on layer 5 is delayed after the second $swap\_stream$ on layer 4. 
Above all, 
since the \added[id=fu]{major} computation happens in Stage II of each thread's sub-domain, 
we avoid the frequent ``layer-wise" thread synchronizations \added[]{that occur} in the wave-front parallelism.
Besides, we only synchronize at the intersection layers every two time steps, 
hence the overhead of three \textit{barriers} of Alg.\ref{alg:3D-2step-prism-omp} 
becomes much less.

\section{Experimental Evaluation}
\label{sec:3D-LBM-exp}



In this section, we first present the \added[]{experimental setup} and validations on our 3D memory-aware LBM.
Then we evaluate its sequential and parallel performance. 
\subsection{Experiment Setup and Verification}
\label{subsec:3D-setup}
The details of our experimental hardware platforms are provided in Table.\ref{tbl:arch}.
To evaluate the performance of our new algorithms,
we use the 3D lid-driven cavity flow \added[]{simulation} as 
\added[]{an example}. 
The 3D cavity has a dimension of $lz \times ly \times lx$, and its top lid moves with a constant velocity $v$.
Our 3D memory-aware LBM algorithms have been implemented as C++ template functions, \added[]{which are then added to the Palabos framework.} 
\added[]{For verification, we construct a $cavity$ with the same procedure, and then separately execute four algorithms on it, i.e.,  Palabos solvers $fuse()$ and $fuse\_prism()$ for $N$ time steps, 
and our memory-aware algorithms $two_\_step\_prism()$ and $two\_step\_prism\_omp()$ for $N/2$ time steps.}
\added[]{Then, we compute the velocity norm of each cell 
and write to four separate logs.
At last, we verify that our algorithms produce the same result} as Palabos
\added[]{for guaranteeing} \added[]{software correctness.}

\begin{table}[h!]
	\renewcommand{\arraystretch}{1.0} 
	\caption{\small Details of our experimental platforms.}
	\centerline{
		{\scriptsize
			\begin{tabularx}{0.883\textwidth}{|c | c | c | c|} 
				\thickhline 
				 & {\it Bridges at PSC} & \multicolumn{2}{c|}{\it Stampede2 at TACC} \\
				\thickhline
				Microarchitecture & {\it Haswell'14} & {\it Skylake'17} & {\it Knight Landing'16}\\
				\hline
				Intel CPU product code & Xeon E5-2695v3 & Xeon Platinum 8160 & Xeon Phi 7250\\
				Total \# Cores/node & 28 on 2 sockets  & 48 on 2 sockets & 68 on 1 socket\\
				Clock rate (GHz) & 2.1$\sim$3.3 & 2.1 nominal(1.4$\sim$3.7) & 1.4\\
				L1 cache/core & 32KB & 32KB &  32KB \\
				L2 cache/core & 256KB & 1MB & 1MB per 2-core tile\\
				L3 cache/socket & 35MB & 33MB (Non-inclusive) & 16GB MCDRAM\\
				DDR4 Memory(GB)/node & 128 (2133 MHz) & 192 (2166 MHz) & 96 (2166 MHz) \\
				\hline
				Compiler & icc/19.5 & \multicolumn{2}{c|}{icc/18.0.2}\\
				AVX extension& AVX2 & \multicolumn{2}{c|}{AVX512}\\
				\thickhline
			\end{tabularx}
		}
	}
	\label{tbl:arch}
\end{table}

\subsection{Performance of Sequential 3D Memory-aware LBM}
\label{subsec:3D-seq}
The first set of experiments with 3D cavity flows compare the sequential performance of four different LBM algorithms, which are the Fuse Swap LBM (with / without prism traversal), and the Two-step Memory-aware LBM (with / without prism traversal).
For simplicity, we use the abbreviations of fuse LBM, fuse prism LBM, 2-step LBM and 2-step prism LBM, respectively.
The problem input are 3D cubes with edge size $L =64 \sim 896$.
\added[id=fu]{Every algorithm with a prism stride configuration is executed five times, and the average MFLUPS (millions of fluid lattice node updates per second) is calculated}.
\added[id=fu, comment={``we use computer time rather than human time to
search a space of code variations for a fixed problem"}]{For the ``prism" algorithms, different prism strides (ranging from 8, 16, 32, ..., to 448) 
are tested, 
and we select the best performance achieved.}

%

\begin{figure}[h!]
     \centering
     \begin{subfigure}[t]{0.32\textwidth}
         \centering
         \includegraphics[width=\textwidth]{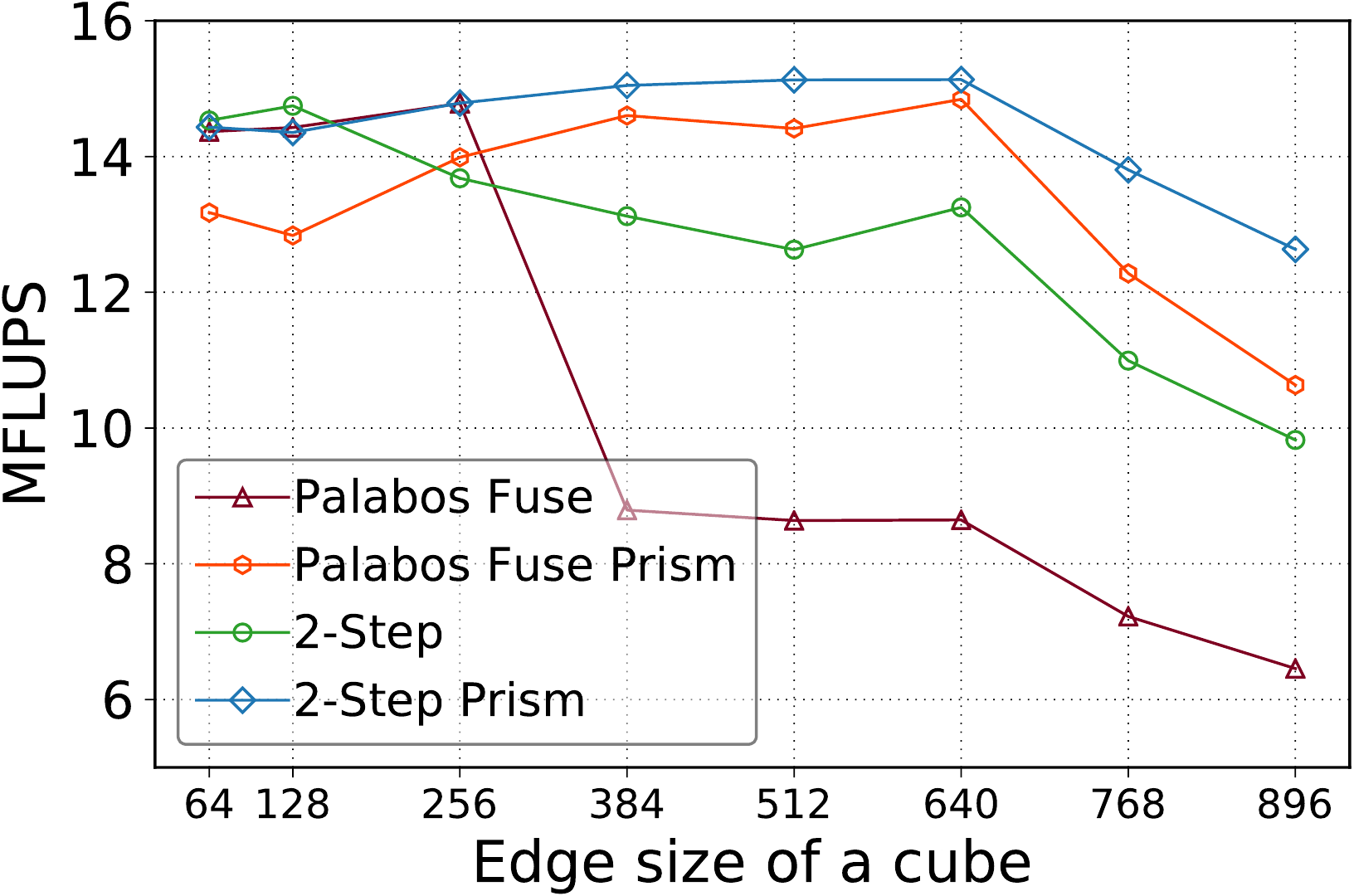}
         \caption{\small Haswell.}
         \label{fig:3D-seq-haswell}
     \end{subfigure}
     \begin{subfigure}[t]{0.32\textwidth}
         \centering
         \includegraphics[width=\textwidth]{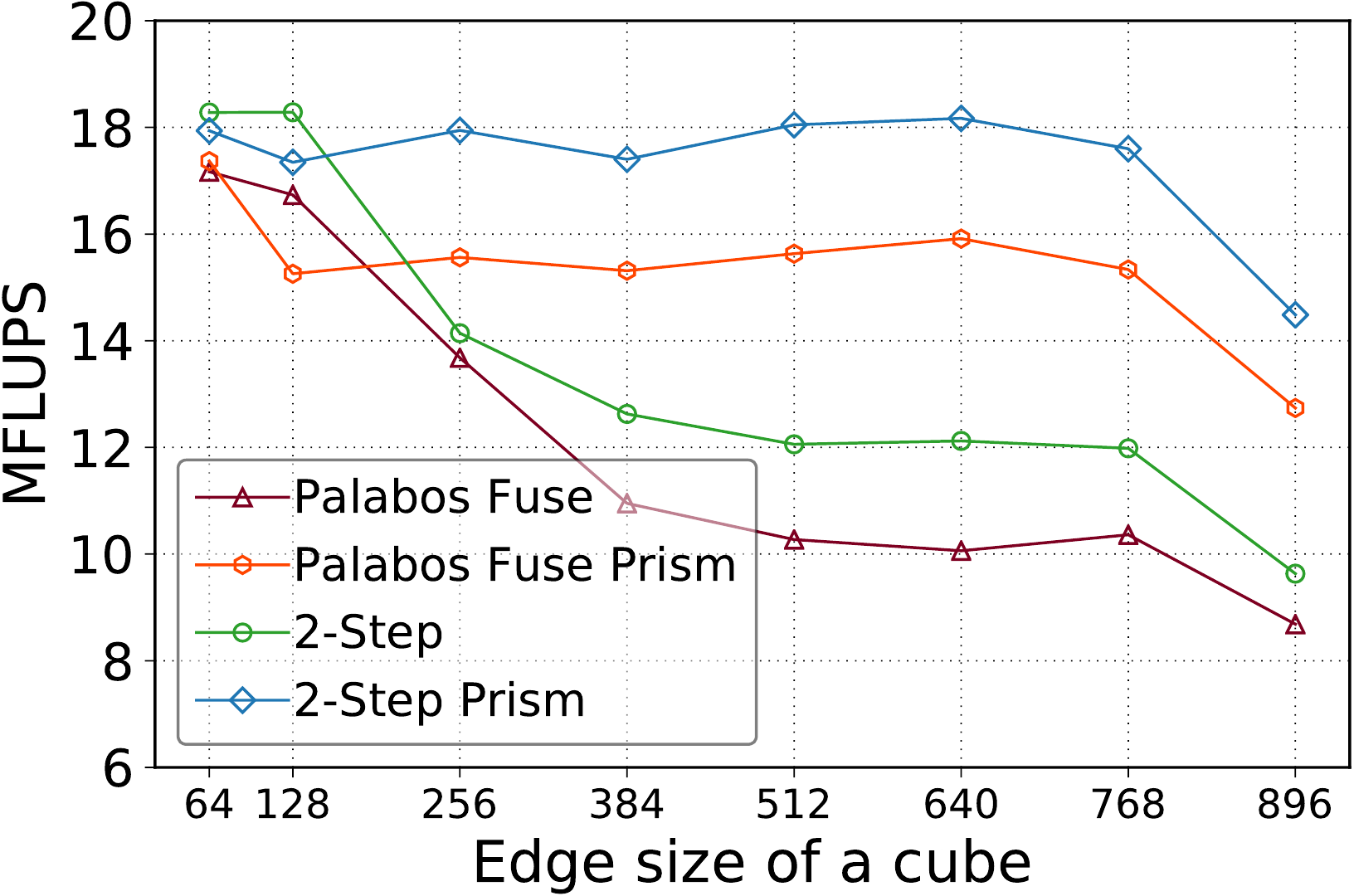}
         \caption{\small Skylake.}
         \label{fig:3D-seq-skx}
     \end{subfigure}
     \begin{subfigure}[t]{0.32\textwidth}
         \centering
         \includegraphics[width=\textwidth]{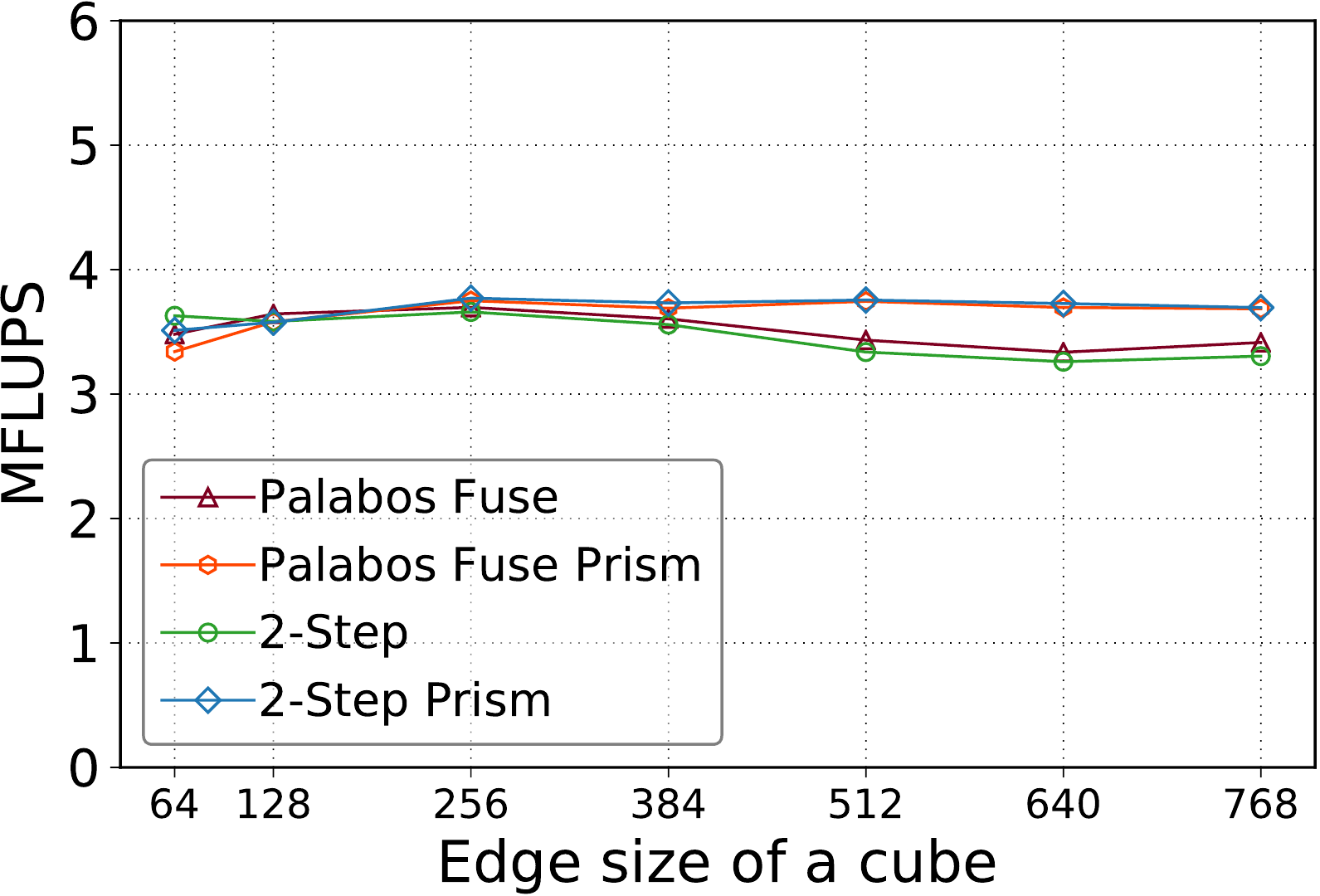}
         \caption{\small Knight Landing.}
         \label{fig:3D-seq-knl}
     \end{subfigure}
    \caption{\small Sequential performance using four LBM algorithms on three types of CPUs.}
    \label{fig:3D-seq}
\end{figure}

Fig.\ref{fig:3D-seq} shows the sequential performance on three types of CPUs. 
When we use small edge sizes (e.g., $L = 64, 128$), 2-step LBM is the fastest. 
But when $L\geq256$, 2-step prism LBM performs the best 
and is up to 18.8\% and 15.5\% faster than the second-fastest Palabos (Fuse Prism LBM solver) on Haswell and Skylake, respectively.
But since KNL does not have an L3 cache, 
2-step prism LBM is only 1.15\% faster than Palabos (Fuse Prism LBM solver). 


We observe that the performance of algorithms without prism traversal starts to drop when $L\geq384$.
Since \added[]{the} swap algorithm streams to half of its neighbors on its own layer and the layer below,
$23.9 MB/layer \times 2 layers = 47.8 MB$ (\added[]{when $L=384$), which exceeds the L3 cache size} (35 MB per socket on Haswell). 
Thus we need to use spatial locality by adding the feature of prism traversal.
Consequently, on Haswell and Skylake,
fuse LBM is improved by up to 71.7\% and 58.2\%, respectively, 
2-step LBM is improved by up to 28.6\% and 50.4\%, respectively.
When only adding the feature of merging two steps, 
2-step LBM is faster than Palabos (Fuse) by up to 53.3\% on Haswell and 20.5\% on Skylake.
Hence, we conclude that both prism traversal and merging two steps significantly increase cache reuse on the large domain.

\begin{figure}[h!]
     \centering
     \begin{subfigure}[t]{0.28\textwidth}
         \centering
         \includegraphics[width=\textwidth]{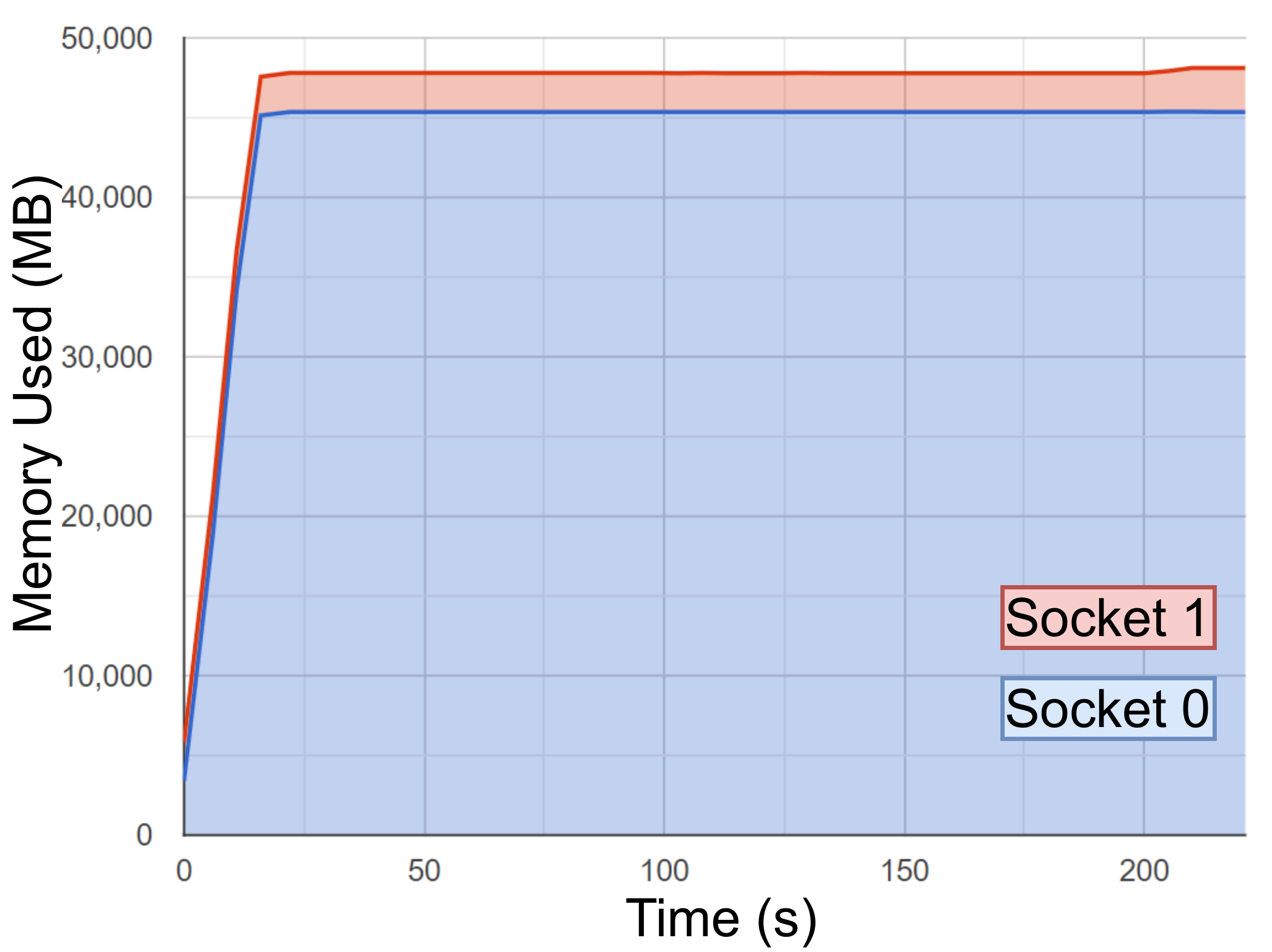}
         \caption{\small $L=640$.} 
         \label{fig:3D-seq-haswell-used-640}
     \end{subfigure}
     \begin{subfigure}[t]{0.28\textwidth}
         \centering
         \includegraphics[width=\textwidth]{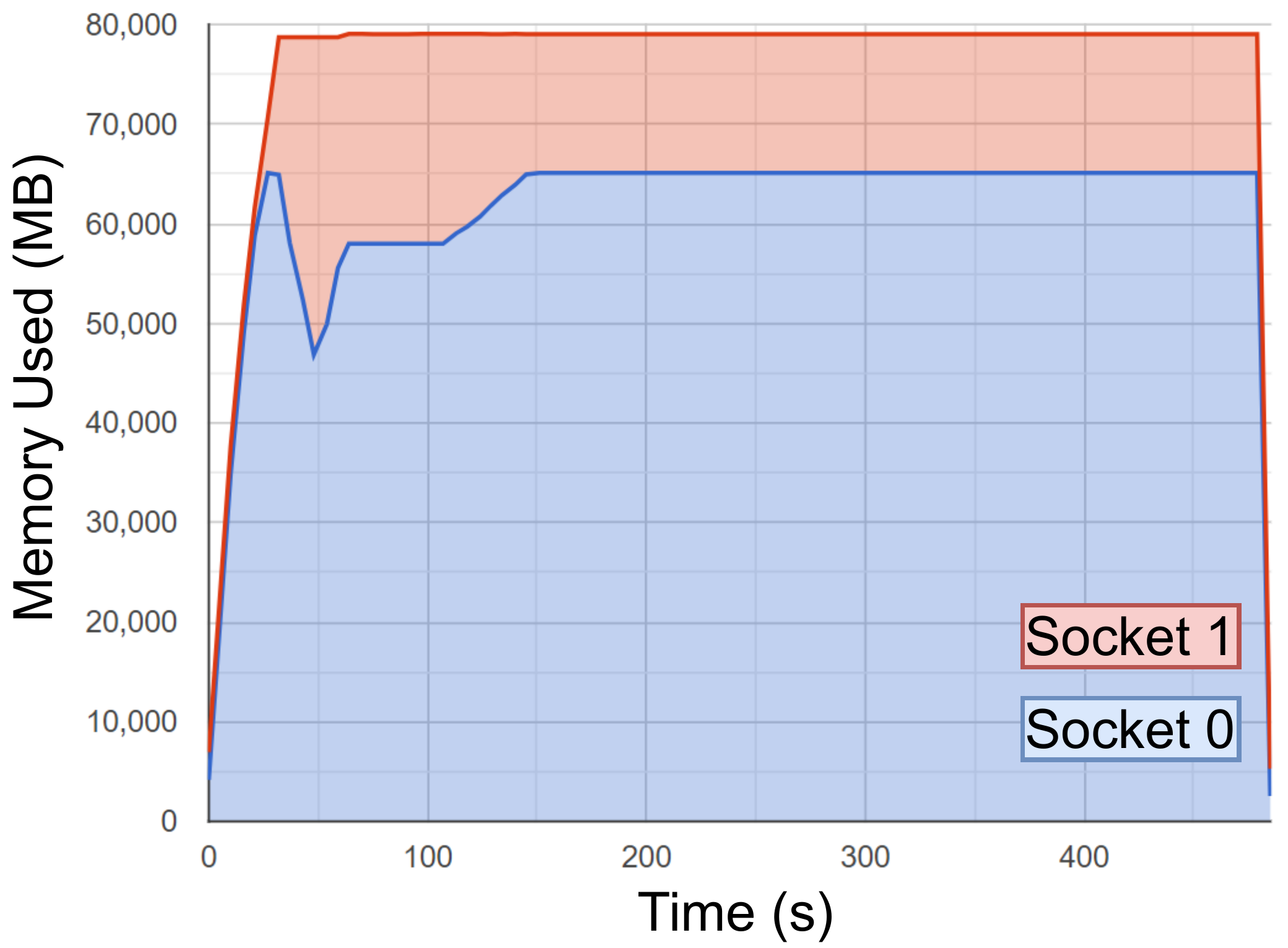}
         \caption{\small $L=768$.}
         \label{fig:3D-seq-haswell-used-768}
     \end{subfigure}
     \begin{subfigure}[t]{0.28\textwidth}
         \centering
         \includegraphics[width=\textwidth]{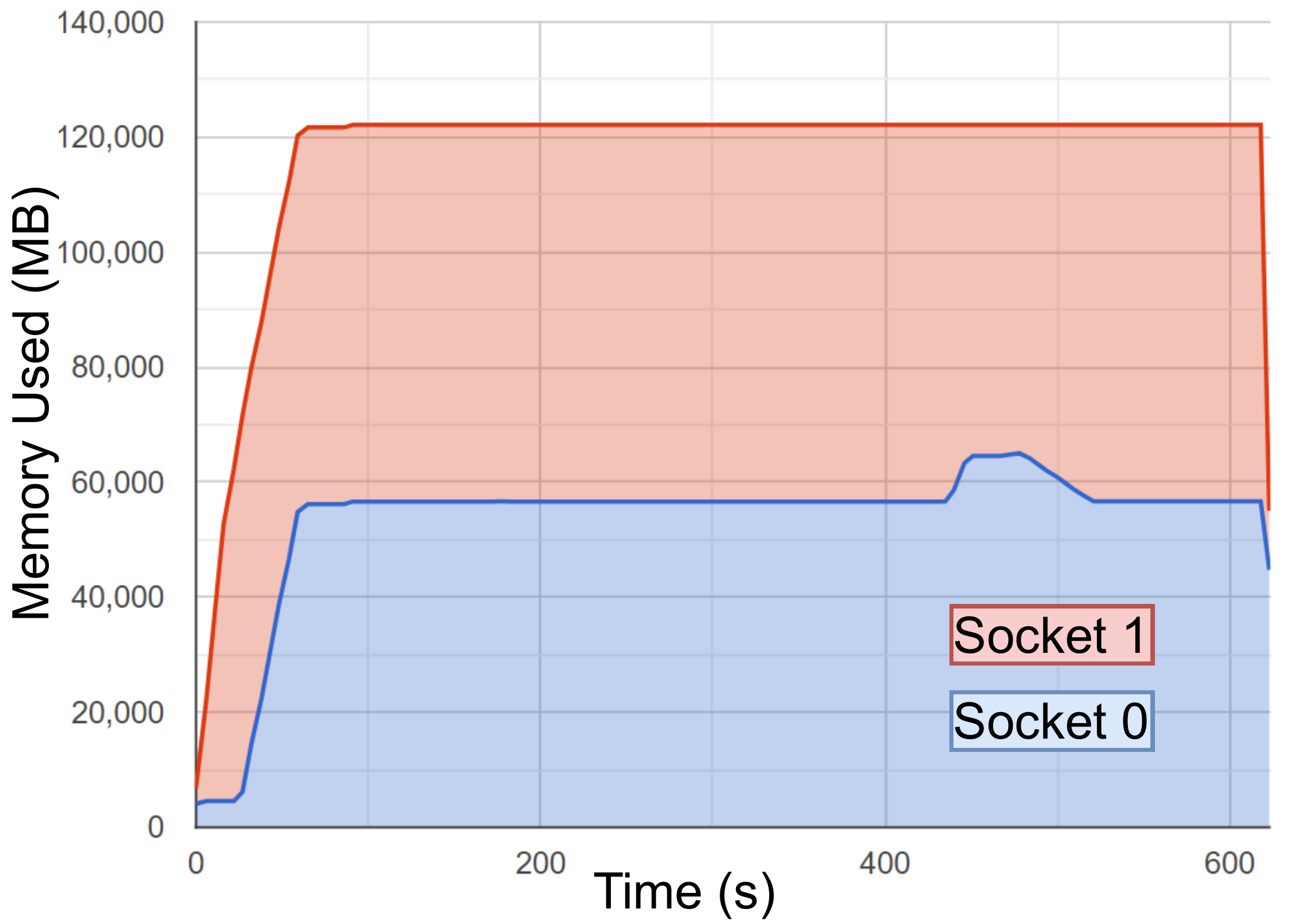}
         \caption{\small $L=896$.}
         \label{fig:3D-seq-haswell-used-896}
     \end{subfigure}
    \caption{Memory usage on two sockets of a Haswell node.}
    \label{fig:3D-seq-numa-haswell}
\end{figure}

In Fig.\ref{fig:3D-seq}, we observe that the performance of all algorithms starts to drop when $L \geq 768$ on Haswell and $L = 896$ on Skylake. 
To find out the reason, 
we use Remora~\cite{rosales2015remora} to monitor the memory usage on each socket of the Haswell node. 
As $L$ increases from 640 to 896, the memory usage on socket 1 (red area) in Fig.\ref{fig:3D-seq-haswell-used-640}$\sim$\ref{fig:3D-seq-haswell-used-896} has enlarged from 2.4 GB to 63.9 GB.
When memory usage exceeds the 64GB DRAM capacity per socket on \added[]{the Haswell node, foreign NUMA memory accesses are involved, thus the sequential performance reduces.} Similar results also happen on the Skylake node. 
However, because the KNL node only has one socket, the performance on KNL does not drop.
\subsection{Performance of Parallel 3D Memory-aware LBM}
\label{subsec:3D-strong-scaling}

Given $N$ cores, Palabos LBM solvers partition the simulation domain evenly along three axes by $N_{z} \times N_{y} \times N_{x} = N$ MPI processes, 
which follows the underlying memory layout of cells along the axis of Z, then Y, and X at last.
But our 3D memory-aware LBM partitions a domain only along X-axis by $N$ OpenMP threads.
Hence, Palabos LBM solvers have a smaller Y-Z layer size per core than our algorithm and have closer memory page alignment especially for a large domain.
To exclude the factor caused by different partition methods, when the input of Palabos LBM solvers still uses cubes, 3D memory-aware LBM will take two different inputs.
Firstly, it takes the input of the ``equivalent dimension" of those cubes, such that
a thread in our algorithm and a process in Palabos will compute a sub-domain with the same dimension after the respective partition method.
Secondly, it simply takes the identical input of those cubes.



\begin{table}[b!] 
	\renewcommand{\arraystretch}{1.0} 
	\caption{\small Equivalent input used by 2-step prism LBM when the input of Palabos LBM solvers is  a cube with $L = 840$ on a Haswell node.}
	\centering
		{\scriptsize
			\begin{tabularx}{0.75\textwidth}{c | c | c | c | c | c | c | c | c |c |c |c} 
				\thickhline 
				Cores & 1	&2	&4	&6	&8	&10	&12	&14	&20	&24	&28\\
				\hline
$lx$ (height) &840	&1680	&3360	&5040	&3360	&8400	&5040	&11760	&8400	&10080	&11760	\\
$ly$ (width)  &840	&840	&420	&420	&420	&420	&420	&420	&420	&420	&420	\\
$lz$ (length) &840	&420	&420	&280	&420	&168	&280	&120	&168	&140	&120	\\
				\thickhline
			\end{tabularx}
		}
	\label{tbl:3D-fair-cube-input-840}
\end{table}

\begin{figure}[t!]
     \centering
     \begin{subfigure}[t]{0.32\textwidth}
         \centering
         \includegraphics[width=\textwidth]{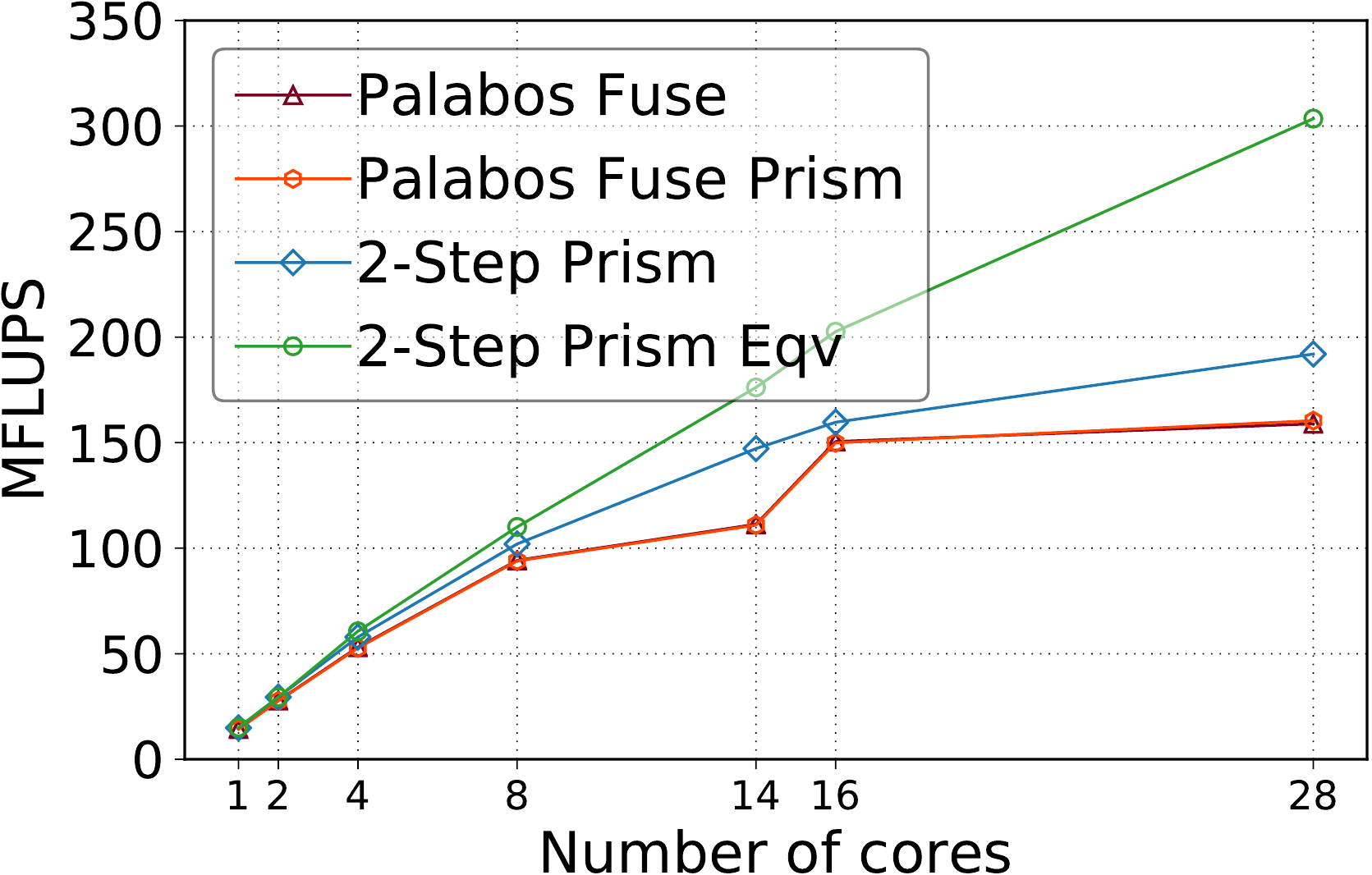}
         \caption{\small Haswell $L = 112$.}
         \label{fig:3D-strong-comb-haswell-112}
     \end{subfigure}
     \begin{subfigure}[t]{0.32\textwidth}
         \centering
         \includegraphics[width=\textwidth]{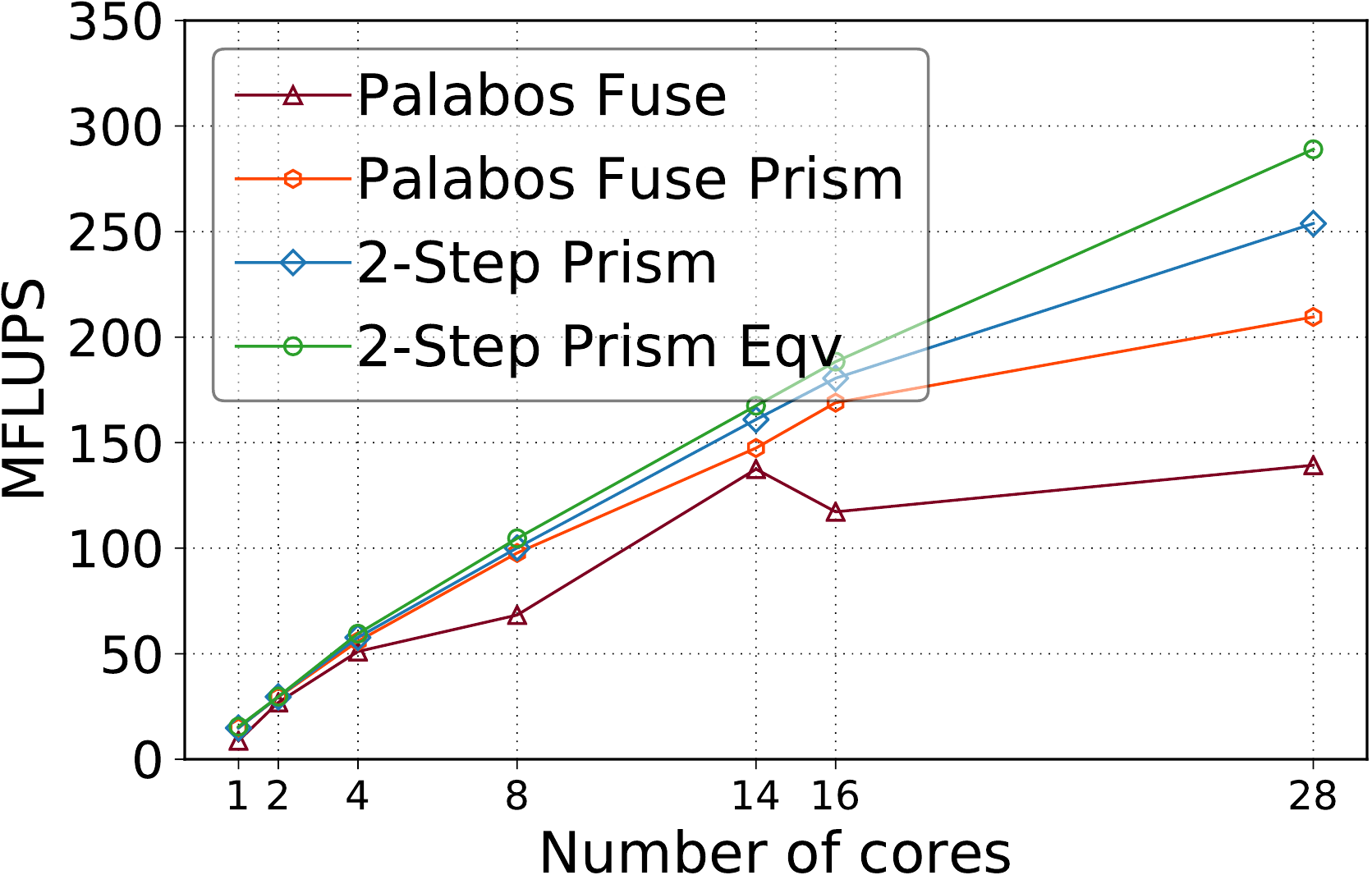}
         \caption{\small Haswell $L = 448$.}
         \label{fig:3D-strong-comb-haswell-448}
     \end{subfigure}
     \begin{subfigure}[t]{0.32\textwidth}
         \centering
         \includegraphics[width=\textwidth]{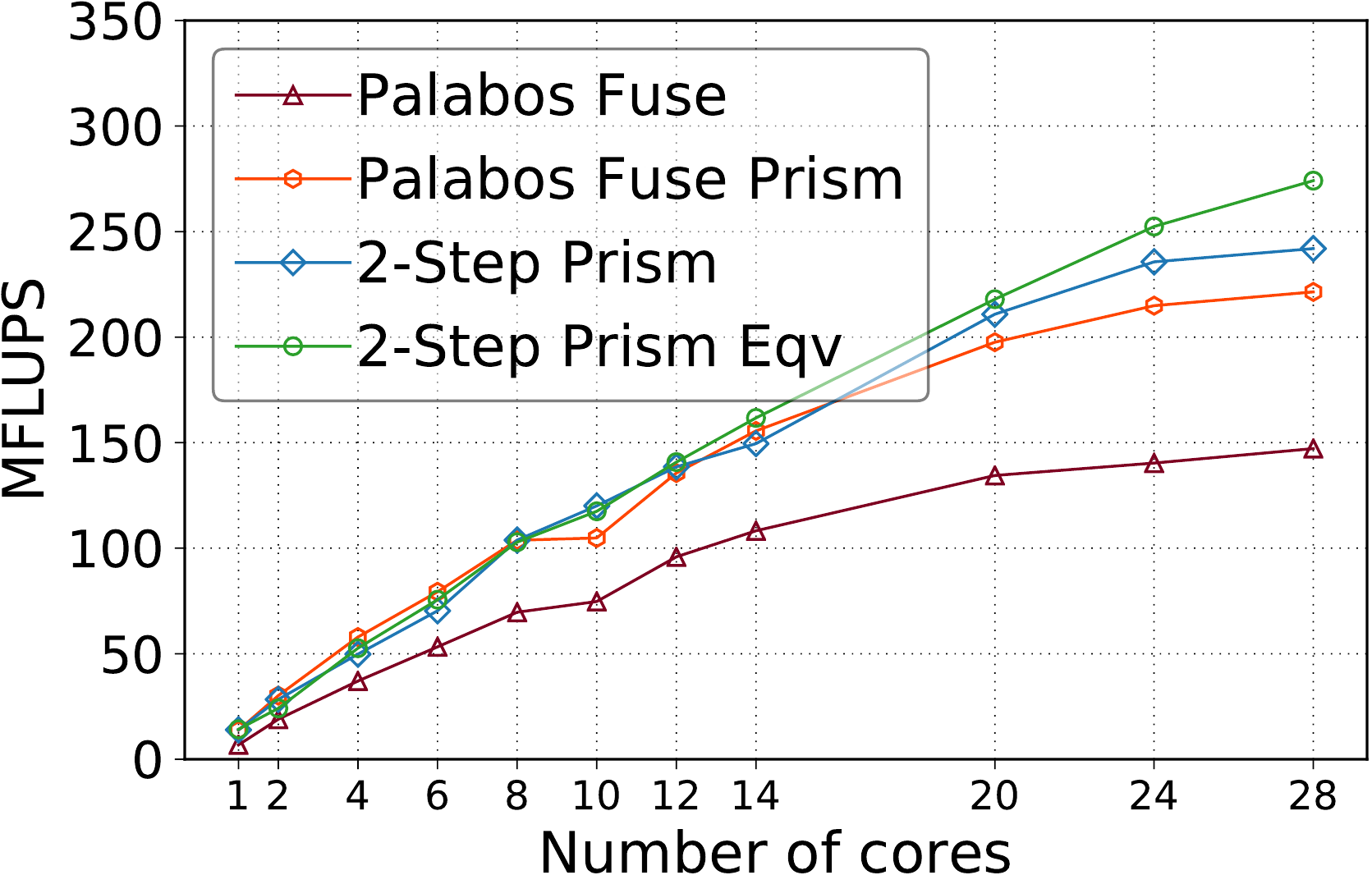}
         \caption{\small Haswell $L = 840$.}
         \label{fig:3D-strong-comb-haswell-840}
     \end{subfigure}
     \begin{subfigure}[t]{0.32\textwidth}
         \centering
         \includegraphics[width=\textwidth]{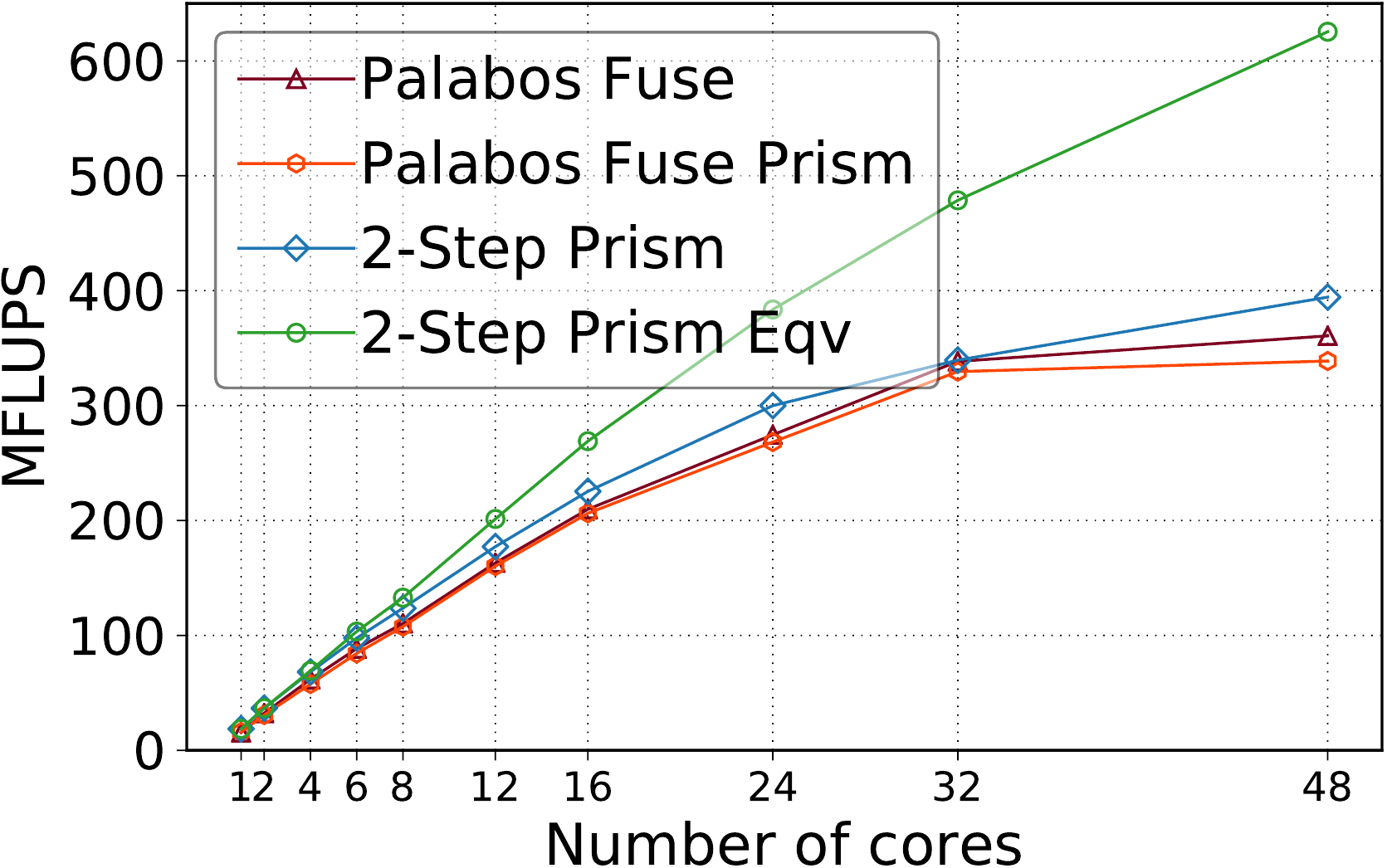}
         \caption{\small Skylake $L = 192$.}
         \label{fig:3D-strong-comb-skx-192}
     \end{subfigure}
     \begin{subfigure}[t]{0.32\textwidth}
         \centering
         \includegraphics[width=\textwidth]{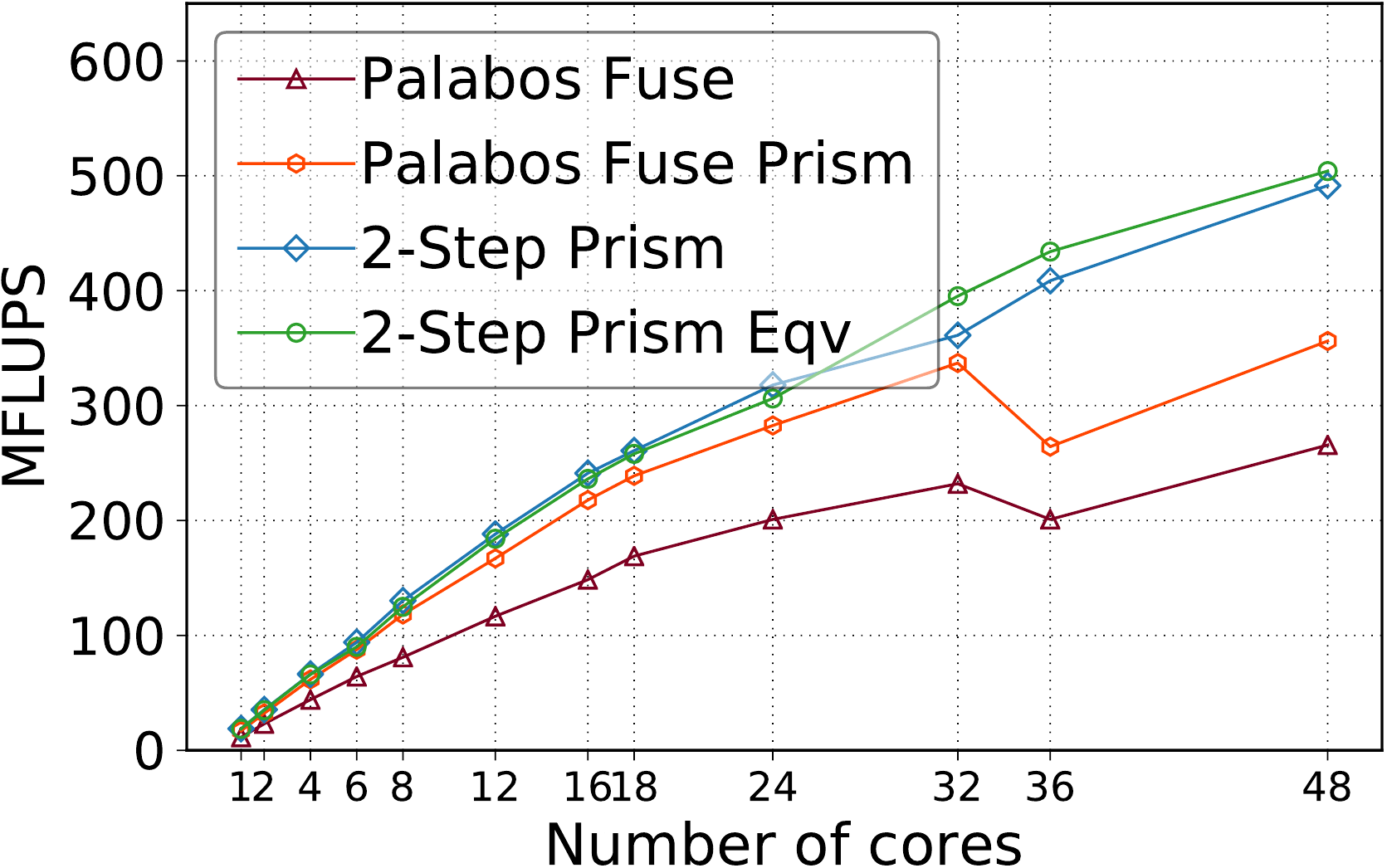}
         \caption{\small Skylake $L = 576$.}
         \label{fig:3D-strong-comb-skx-576}
     \end{subfigure}
     \begin{subfigure}[t]{0.32\textwidth}
         \centering
         \includegraphics[width=\textwidth]{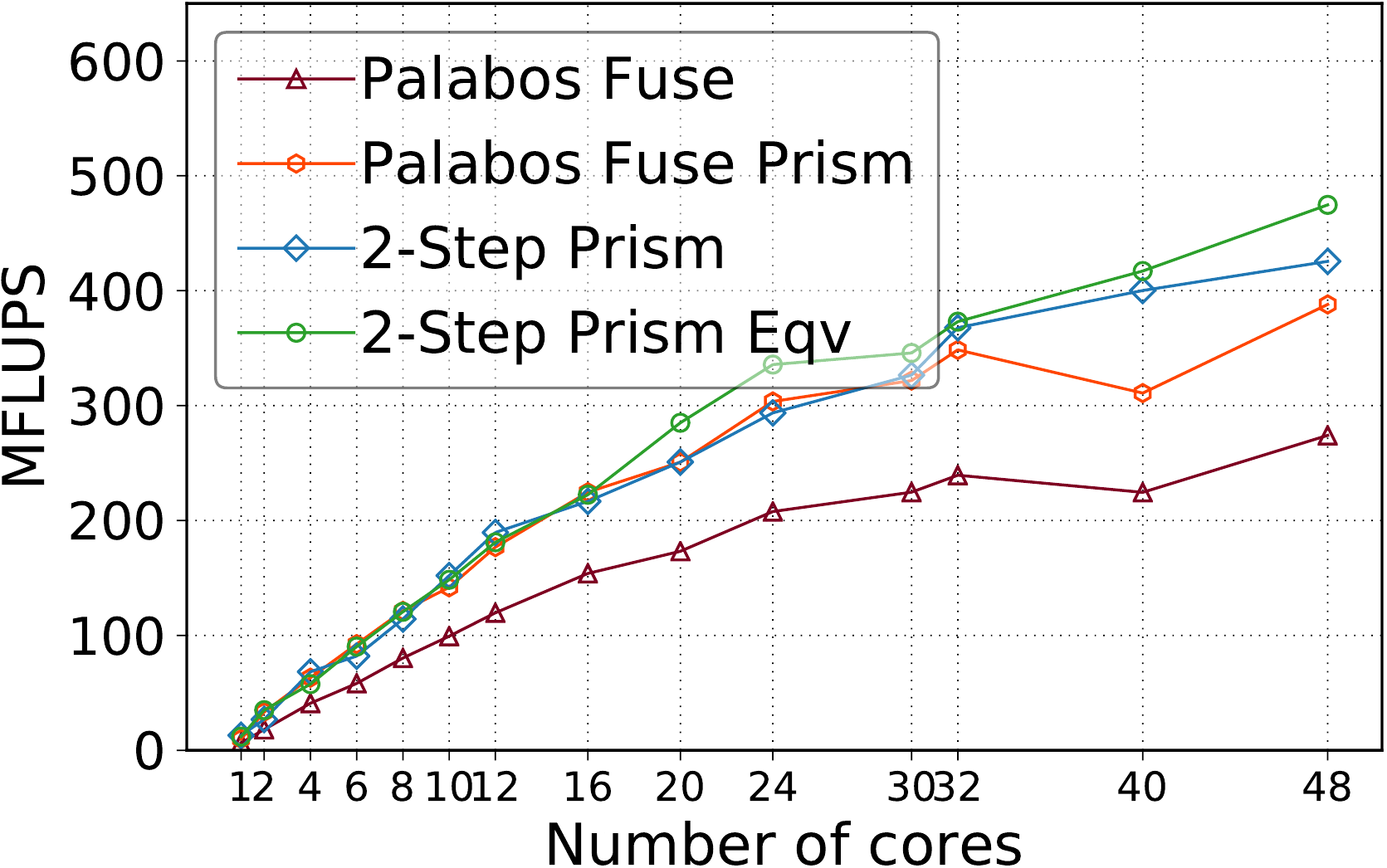}
         \caption{\small Skylake $L = 960$.}
         \label{fig:3D-strong-comb-skx-960}
     \end{subfigure}
     \begin{subfigure}[t]{0.32\textwidth}
         \centering
         \includegraphics[width=\textwidth]{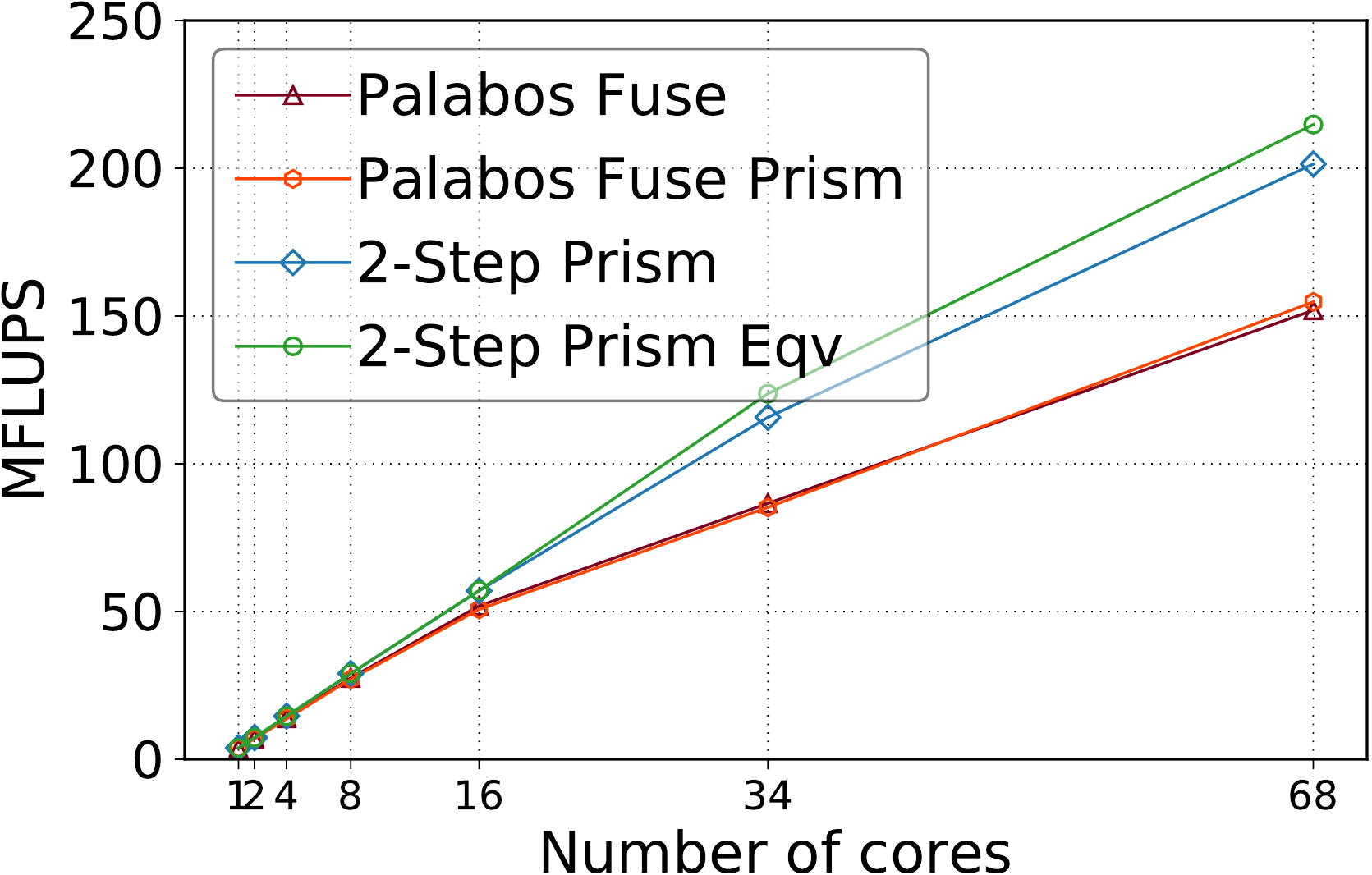}
         \caption{\small KNL $L=272$.}
         \label{fig:3D-strong-comb-knl-272}%
     \end{subfigure}
     \begin{subfigure}[t]{0.32\textwidth}
         \centering
         \includegraphics[width=\textwidth]{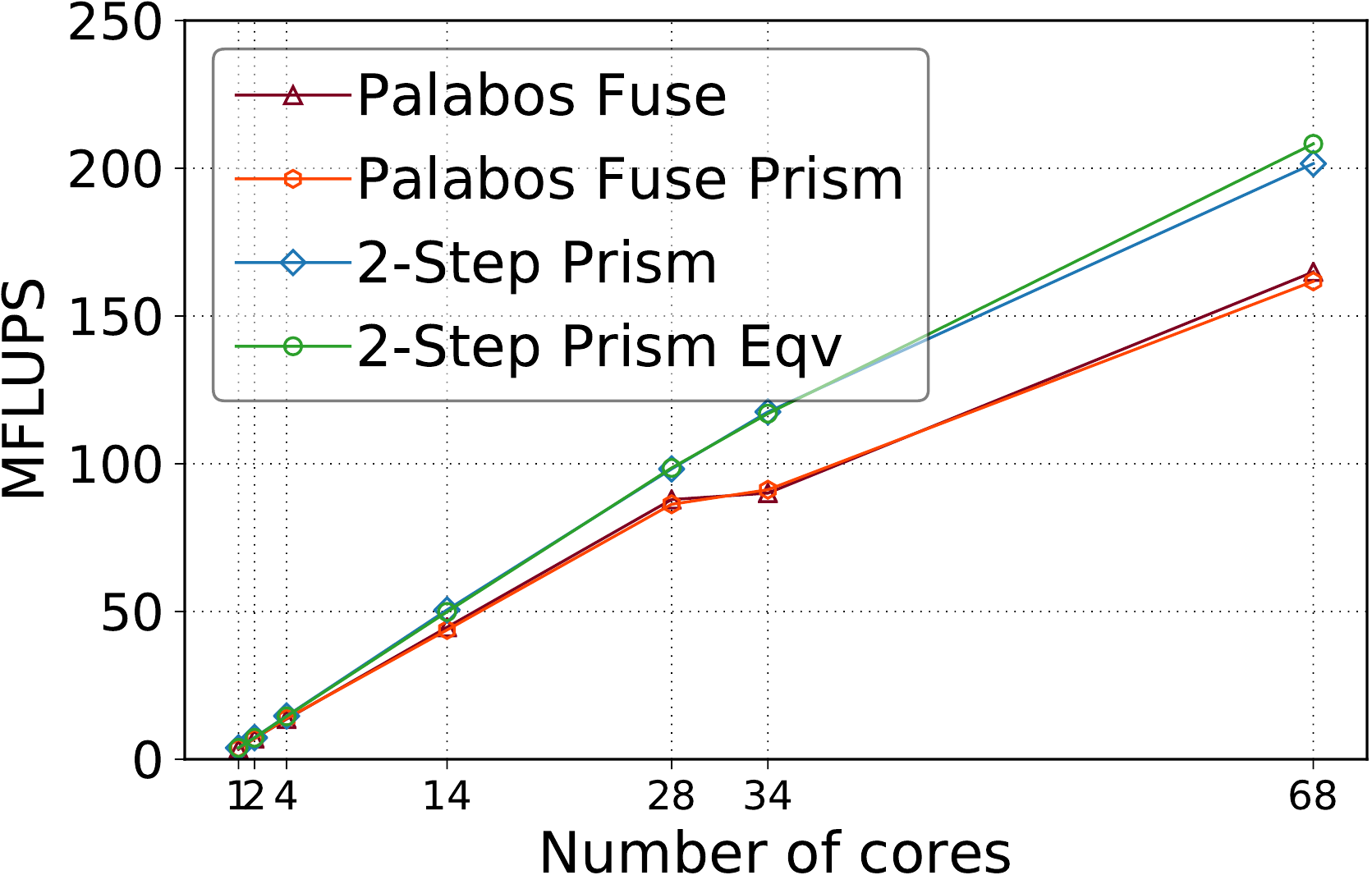}
         \caption{\small KNL $L=476$.}
         \label{fig:3D-strong-comb-knl-476}%
     \end{subfigure}
     \begin{subfigure}[t]{0.32\textwidth}
         \centering
         \includegraphics[width=\textwidth]{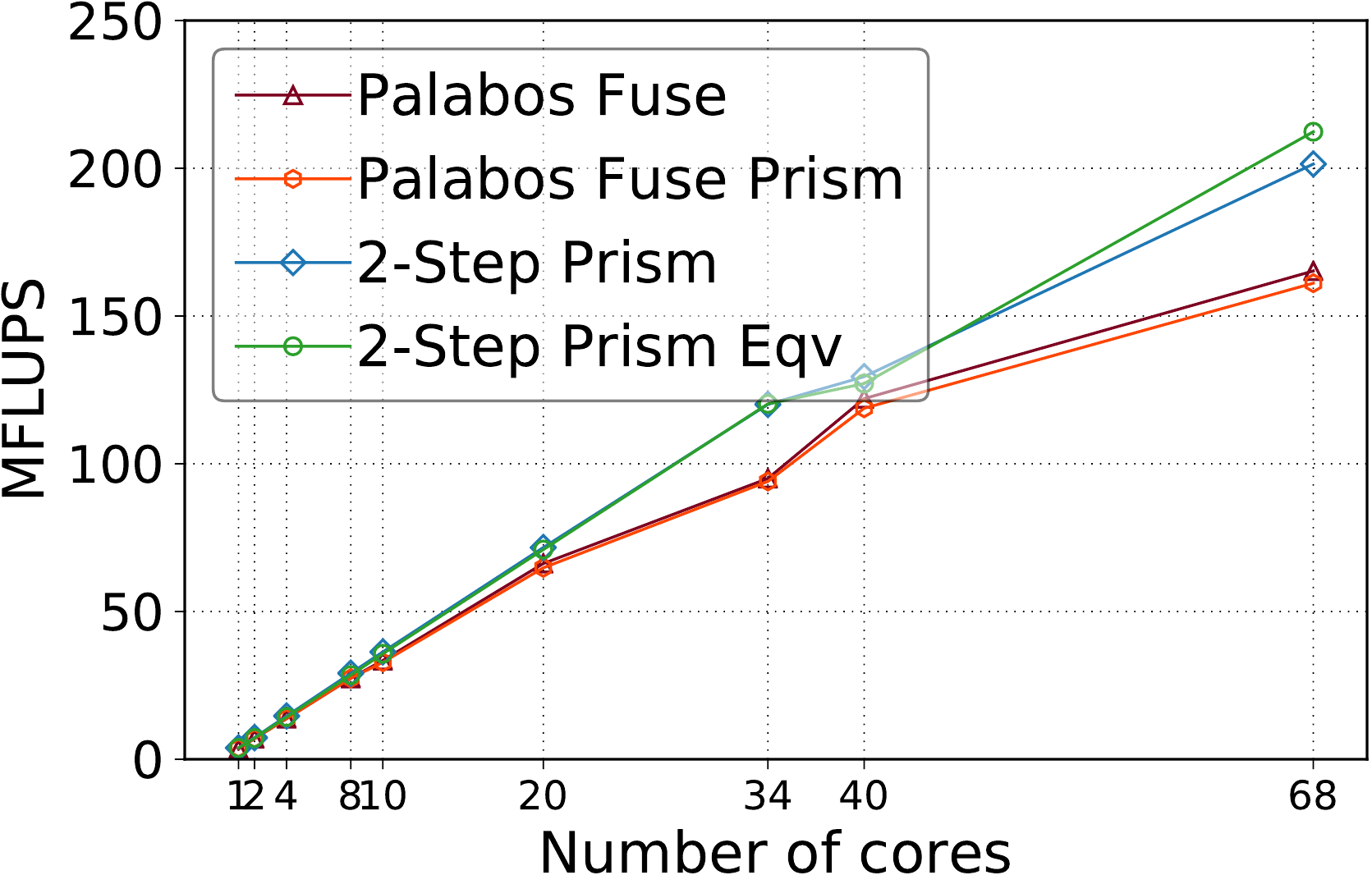}
         \caption{\small KNL $L=680$.}
         \label{fig:3D-strong-comb-knl-680}%
     \end{subfigure}
    
    \caption{\small Strong scalability performance on three types of compute nodes. 
    ``2-step prism eqv" = Parallel 3D memory aware LBM takes the equivalent input of cubes.}
    \label{fig:3D-strong-comb-cube}%
\end{figure}

Fig.\ref{fig:3D-strong-comb-cube} shows the strong scalability of three LBM algorithms on three types of compute nodes. The input of Palabos LBM solvers use cubes with edge size $L$ from small to large. 
Tab.\ref{tbl:3D-fair-cube-input-840} gives an example of the equivalent input used by 3D memory-aware LBM 
when Palabos LBM solvers use a cube with $L = 840$ on a Haswell node.
We observe that the 2-step prism LBM scales efficiently and always achieves the best performance in all cases. 
(1) When using the equivalent input of cubes,
for small scale cubes (with $L=112, 192, 272$) in Fig.\ref{fig:3D-strong-comb-haswell-112}.\ref{fig:3D-strong-comb-skx-192}.\ref{fig:3D-strong-comb-knl-272},  
3D memory-aware LBM (green legend) is faster than the second-fastest Palabos (Fuse Prism) (orange legend) by up to 89.2\%, 84.6\%, and 38.8\% \added[id=fu]{on the Haswell, Skylake, and KNL node, respectively.
Missing L3 cache on KNL prevents the similar speedup as other two CPUs.}
In Fig.\ref{fig:3D-strong-comb-haswell-448}.\ref{fig:3D-strong-comb-skx-576}.\ref{fig:3D-strong-comb-knl-476}, for the middle scale cubes (with $L=448, 576, 476$),
it is still faster than Palabos (Fuse Prism) by up to 37.9\%, 64.2\%, and 28.8\% on three CPU nodes, respectively.
\added[id=fu]{Due to unbalanced number of processes assigned on three axes, 
we observe that the performance of Palabos Fuse and Fuse Prism drop on some number of cores.}
In Fig.\ref{fig:3D-strong-comb-haswell-840}.\ref{fig:3D-strong-comb-skx-960}.\ref{fig:3D-strong-comb-knl-680}, for the large scale cubes (with $L=840, 960, 680$), 
it is still faster than Palabos (Fuse Prism) by up to 34.2\%, 34.2\%, and 31.8\%, respectively. 
(2) When using the identical input of cubes, although our 3D memory-aware LBM has larger Y-Z layer sizes, 
it is still faster than Palabos (Fuse Prism) but with less speedup than before, i.e., by up to 21.1\%, 54.7\%, and 30.1\% on three CPU nodes, respectively.
\added[id=fu]{The less speedup suggests our future work to partition a 3D domain along three axes to utilize closer memory page alignment on smaller Y-Z layer size.}

\section{Conclusion}

To address the memory-bound limitation of LBM, 
we design \added[]{a new 3D parallel memory-aware LBM algorithm} that systematically combines single copy distribution, single sweep, swap algorithm, prism traversal, and merging two collision-streaming cycles.
We also keep thread safety and reduce the synchronization cost in parallel.
The parallel 3D memory-aware LBM outperforms state-of-the-art LBM software by up to 89.2\% on a Haswell node, 84.6\% on a Skylake node and 38.8\% on a Knight Landing node, respectively.
Our future work is to  
\added[id=fu]{merge more time steps on distributed memory systems and on GPU}.








\bibliographystyle{splncs04}
\bibliography{all}

\begin{thebibliography}{10}
\providecommand{\url}[1]{\texttt{#1}}
\providecommand{\urlprefix}{URL }
\providecommand{\doi}[1]{https://doi.org/#1}

\bibitem{LBM-AA}
Bailey, P., Myre, J., Walsh, S.D., Lilja, D.J., Saar, M.O.: {Accelerating
  lattice Boltzmann fluid flow simulations using graphics processors}. In: 2009
  international conference on parallel processing. pp. 550--557. IEEE (2009)

\bibitem{coreixas2019comprehensive}
Coreixas, C., Chopard, B., Latt, J.: Comprehensive comparison of collision
  models in the lattice {Boltzmann} framework: Theoretical investigations.
  Physical Review E  \textbf{100}(3),  033305 (2019)

\bibitem{crimi2013early}
Crimi, G., Mantovani, F., Pivanti, M., Schifano, S.F., Tripiccione, R.: {Early
  experience on porting and running a Lattice Boltzmann code on the Xeon-Phi
  co-processor}. Procedia Computer Science  \textbf{18},  551--560 (2013)

\bibitem{waLBerla-paper}
Feichtinger, C., Donath, S., K{\"o}stler, H., G{\"o}tz, J., R{\"u}de, U.:
  Walberla: Hpc software design for computational engineering simulations.
  Journal of Computational Science  \textbf{2}(2),  105--112 (2011)

\bibitem{fu2018designing}
Fu, Y., Li, F., Song, F., Zhu, L.: Designing a parallel memory-aware lattice
  {Boltzmann} algorithm on manycore systems. In: 30th International Symposium
  on Computer Architecture and High Performance Computing. pp. 97--106. IEEE
  (2018)

\bibitem{LBM-esoteric-twist}
Geier, M., Sch{\"o}nherr, M.: Esoteric twist: an efficient in-place streaming
  algorithms for the lattice {Boltzmann} method on massively parallel hardware.
  Computation  \textbf{5}(2), ~19 (2017)

\bibitem{habich2009enabling}
Habich, J., Zeiser, T., Hager, G., Wellein, G.: {Enabling temporal blocking for
  a lattice Boltzmann flow solver through multicore-aware wavefront
  parallelization}. In: 21st International Conference on Parallel Computational
  Fluid Dynamics. pp. 178--182 (2009)

\bibitem{heuveline2007openlb}
Heuveline, V., Latt, J.: The openlb project: an open source and object oriented
  implementation of lattice boltzmann methods. International Journal of Modern
  Physics C  \textbf{18}(04),  627--634 (2007)

\bibitem{latt2007technical}
Latt, J.: Technical report: How to implement your ddqq dynamics with only q
  variables per node (instead of 2q). Tufts University pp.~1--8 (2007)

\bibitem{latt2020palabos}
Latt, J., Malaspinas, O., Kontaxakis, D., Parmigiani, A., Lagrava, D., Brogi,
  F., Belgacem, M.B., Thorimbert, Y., Leclaire, S., Li, S., et~al.: {Palabos:
  Parallel Lattice {Boltzmann} solver}. Computers \& Mathematics with
  Applications  (2020)

\bibitem{liu2017accelerating}
Liu, S., Zou, N., et~al.: Accelerating the parallelization of lattice
  {Boltzmann} method by exploiting the temporal locality. In: International
  Symposium on Parallel and Distributed Processing with Applications. pp.
  1186--1193. IEEE (2017)

\bibitem{malas2015multicore}
Malas, T., Hager, G., Ltaief, H., Stengel, H., Wellein, G., Keyes, D.:
  Multicore-optimized wavefront diamond blocking for optimizing stencil
  updates. SIAM Journal on Scientific Computing  \textbf{37}(4),  C439--C464
  (2015)

\bibitem{LBM-swap}
Mattila, K., Hyv{\"a}luoma, J., Rossi, T., Aspn{\"a}s, M., Westerholm, J.: An
  efficient swap algorithm for the lattice {Boltzmann} method. Computer Physics
  Communications  \textbf{176}(3),  200--210 (2007)

\bibitem{Hemelb-paper}
Mazzeo, M.D., Coveney, P.V.: {HemeLB: A high performance parallel
  lattice-{Boltzmann} code for large scale fluid flow in complex geometries}.
  Computer Physics Communications  \textbf{178}(12),  894--914 (2008)

\bibitem{Musubi}
{Musubi}: \url{https://geb.sts.nt.uni-siegen.de/doxy/musubi/index.html} (2021)

\bibitem{OPENMP}
{OpenMP}: \url{http://www.openmp.org} (2021)

\bibitem{Palabos}
{Palabos}: \url{https://palabos.unige.ch/} (2021)

\bibitem{perepelkina2018lrnla}
Perepelkina, A., Levchenko, V.: {LRnLA algorithm ConeFold with non-local
  vectorization for LBM implementation}. In: Russian Supercomputing Days. pp.
  101--113. Springer (2018)

\bibitem{LBM-shift}
Pohl, T., Kowarschik, M., Wilke, J., Iglberger, K., R{\"u}de, U.: {Optimization
  and profiling of the cache performance of parallel lattice Boltzmann codes}.
  Parallel Processing Letters  \textbf{13}(04),  549--560 (2003)

\bibitem{randles2013performance}
Randles, A.P., Kale, V., Hammond, J., Gropp, W., Kaxiras, E.: Performance
  analysis of the lattice {Boltzmann} model beyond navier-stokes. In: 27th
  International Symposium on Parallel and Distributed Processing. pp.
  1063--1074. IEEE (2013)

\bibitem{rivera2000tiling}
Rivera, G., Tseng, C.W.: Tiling optimizations for 3d scientific computations.
  In: SC'00: Proceedings of the 2000 ACM/IEEE conference on Supercomputing. pp.
  32--32. IEEE (2000)

\bibitem{rosales2015remora}
Rosales, C., etc.: Remora: a resource monitoring tool for everyone. In:
  Proceedings of the Second International Workshop on HPC User Support Tools.
  pp.~1--8 (2015)

\bibitem{slaughter2020task}
Slaughter, E., Wu, W., Fu, Y., Brandenburg, L., Garcia, N., Kautz, W., Marx,
  E., Morris, K.S., , et~al.: {Task bench: A parameterized benchmark for
  evaluating parallel runtime performance}. In: SC'20: International Conference
  for High Performance Computing, Networking, Storage and Analysis. pp. 1--15.
  IEEE (2020)

\bibitem{succi2019towards}
Succi, S., Amati, G., Bernaschi, M., Falcucci, G., et~al.: Towards exascale
  lattice {Boltzmann} computing. Computers \& Fluids  \textbf{181},  107--115
  (2019)

\bibitem{valero2017reducing}
Valero-Lara, P.: {Reducing memory requirements for large size LBM simulations
  on GPUs}. Concurrency \& Computation: Practice \& Experience
  \textbf{29}(24),  e4221 (2017)

\bibitem{vardhan2019moment}
Vardhan, M., Gounley, J., Hegele, L., Draeger, E.W., Randles, A.: Moment
  representation in the lattice {Boltzmann} method on massively parallel
  hardware. In: SC'19: Proceedings of the International Conference for High
  Performance Computing, Networking, Storage and Analysis. pp. 1--21 (2019)

\bibitem{wellein2009efficient}
Wellein, G., Hager, G., Zeiser, T., Wittmann, M., Fehske, H.: {Efficient
  temporal blocking for stencil computations by multicore-aware wavefront
  parallelization}. In: 33rd Annual IEEE International Computer Software and
  Applications Conference. vol.~1, pp. 579--586. IEEE (2009)

\bibitem{witherden2017future}
Witherden, F.D., Jameson, A.: Future directions in computational fluid
  dynamics. In: 23rd AIAA Computational Fluid Dynamics Conference. p.~3791
  (2017)

\bibitem{wittmann2013comparison}
Wittmann, M., Zeiser, T., Hager, G., Wellein, G.: {Comparison of different
  propagation steps for lattice Boltzmann methods}. Computers \& Mathematics
  with Applications  \textbf{65}(6),  924--935 (2013)

\bibitem{Hemocell-paper}
Zavodszky, G., van Rooij, B., Azizi, V., Alowayyed, S., Hoekstra, A.: Hemocell:
  a high-performance microscopic cellular library. Procedia Computer Science
  \textbf{108},  159--165 (2017)

\bibitem{zeiser2008introducing}
Zeiser, T., Wellein, G., Nitsure, A., Iglberger, K., Rude, U., Hager, G.:
  {Introducing a parallel cache oblivious blocking approach for the lattice
  Boltzmann method}. Progress in Computational Fluid Dynamics, an International
  Journal  \textbf{8}(1-4),  179--188 (2008)

\end{thebibliography}

\end{document}